\documentclass[%
 reprint, %linenumbers,
superscriptaddress,
nofootinbib,
 amsmath,amssymb,
 aps, physrev,
]{revtex4-2}
\usepackage{placeins}
\usepackage{graphicx}% Include figure files
\usepackage{dcolumn}% Align table columns on decimal point
\usepackage{subcaption}
\usepackage{bm}% bold math
\usepackage{hyperref}% add hypertext capabilities
\usepackage{float}
\usepackage{xcolor}
\usepackage{makecell}
\hypersetup{
    colorlinks=true, %set true if you want colored links
    linktoc=all,     %set to all if you want both sections and subsections linked
    linkcolor=blue,  %choose some color if you want links to stand out
    breaklinks=true
}

\newcommand{\vars}{\bm{\alpha}}

\begin{document}

\preprint{APS/123-QED}

\title{Bayesian calibration of a regional optical potential and uncertainty-quantified predictions for compound nucleus reactions}

\author{S. Sullivan}
\email{samuel.sullivan@surrey.ac.uk}
\affiliation{
 School of Mathematics and Physics, University of Surrey, Guildford, Surrey, GU2 7XH, UK
}
\affiliation{Facility for Rare Isotope Beams, Michigan State University, East Lansing, Michigan, 48824, USA}

\author{K. Beyer}
\email{beyerk@frib.msu.edu}
\affiliation{Facility for Rare Isotope Beams, Michigan State University, East Lansing, Michigan, 48824, USA}

\author{F.M. Nunes}
\email{nunes@frib.msu.edu}
\affiliation{Facility for Rare Isotope Beams, Michigan State University, East Lansing, Michigan, 48824, USA}
\affiliation{Department of Physics and Astronomy, Michigan State University, East Lansing, Michigan, 48824, USA}

\author{P. Stevenson}
\affiliation{
 School of Mathematics and Physics, University of Surrey, Guildford, Surrey, GU2 7XH, UK
}
\affiliation{
 AWE Nuclear Security Technologies, Aldermaston, Berkshire, RG7 4PR, UK
}

\author{J. Benstead}
\affiliation{
 AWE Nuclear Security Technologies, Aldermaston, Berkshire, RG7 4PR, UK
}
\affiliation{
 School of Mathematics and Physics, University of Surrey, Guildford, Surrey, GU2 7XH, UK
}

\author{L. Morgan}
\affiliation{
 AWE Nuclear Security Technologies, Aldermaston, Berkshire, RG7 4PR, UK
}
\affiliation{
 School of Mathematics and Physics, University of Surrey, Guildford, Surrey, GU2 7XH, UK
}

\begin{abstract}
\begin{center}
\begin{minipage}{0.85\textwidth}
\begin{description}
\item[Background] Optical potentials are widely used in nuclear physics. Uncertainty-quantified global parameterizations are useful for making systematic studies across the nuclear chart and making predictions away from stability. While they generally provide a good description of the reactions on which they were calibrated, predictions can deviate strongly in specific cases and extrapolations incur large uncertainties. 
\item[Purpose] We propose a framework for a regional optical potential, suited for a specific region of the nuclear landscape. Our goal is to calibrate the potential along isotopic chains to enable both strong data coverage and improved extrapolative power to neighboring isotopes. 
\item[Method] We modify the Lane-consistent Chapel Hill optical potential and the statistical model used previously by Pruitt \textit{et al.} to perform a Bayesian analysis of elastic scattering of neutrons and protons from zirconium isotopes with improved data coverage. The parameter distributions of our Chapel Hill Regional Potential (CHiRP) are propagated to compound nucleus reaction observables including $(n,n')$, $(n,\gamma)$ and $(n,2n)$, which have relevance to various nuclear technology applications. Results are compared to CHUQ, a global calibration of the Chapel Hill potential, published by Pruitt \textit{et al.}.
\item[Results] The final parameterization of CHiRP differs from CHUQ particularly in the energy and radial dependence of its imaginary components. For elastic scattering, the uncertainties on the observables using CHiRP are smaller than those using CHUQ. CHiRP also provides an improvement over its global counterpart in the agreement with data around and above $E \approx 10$ MeV, while the regional and global approaches offer similar levels of agreement to the higher energy and heavier target calibration data. The relative performance of CHiRP and CHUQ when propagating their uncertainty through compound nucleus reaction channels is also reported.
\item[Conclusion] The regional potential approach is a viable alternative to global optical model parameterizations for applications that require precise information within a sub-region of the nuclear landscape.
\end{description}
\end{minipage}
\end{center}
\end{abstract}

\maketitle

\section{\label{sec:intro}Introduction}

Neutron-induced fission has great relevance to society \cite{lrp2023}. Fission applications extend from the nuclear energy sector, to nuclear astrophysics, to national defense programs \cite{jenkins2025}. The fission of actinides populates a range of lighter isotopes that may subsequently undergo reactions with neutrons. When modeling these complex reaction networks, it is essential to know the cross sections of all relevant neutron-induced channels. In addition to the elastic channel, nuclear data is needed for e.g. $(n,n')$, $(n,\gamma)$, $(n,2n)$, $(n,p)$, etc., over a wide range of neutron energies \cite{jenkins2025}. While some experimental data are available, the nuclear data community relies on reaction models to determine the missing pieces. These reaction models are often simulated using reaction codes such as {\sc talys} \cite{koning2023}, {\sc empire} \cite{herman2007}, and {\sc yahfc} \cite{ormand2021}. Fragments populated through fission include neutron-rich isotopes away from stability for which there are no experimental data \cite{beyer2024}. Consequently, models must reliably extrapolate within the relevant region and provide a useful estimate of the uncertainty when doing so.

The reactions mentioned above occur primarily as compound processes, whereby a neutron is captured by the fission fragment of mass $A$, forming a compound state in the $A+1$ system, which then de-excites, evolving towards the final output channel. One important ingredient in modeling these compound processes is the neutron-target effective interaction, usually described by an optical model potential (OMP) \cite{op2023}. The optical potential is rooted in microscopic theory and a wide variety of such theories have been developed to extract it \cite{op2023}. These range from the most fundamental theories (e.g. \cite{rotureau2017,idini2019,vorabbi_microscopic_2024}) to density functional theories (e.g. \cite{bauge2001,blanchon2014}). However, for practical purposes and applications, calibration to experimental data is needed. Calibrations  primarily use elastic scattering measurements. In addition, the dispersive relation can further constrain the optical potential, with the advantage that it provides a natural connection between bound and scattering states \cite{mahzoon2014,mahzoon2017,atkinson2020,pruitt2020}.

A large body of work has been published on the understanding and characterization of uncertainties in local OMPs using Bayesian analyses \cite{lovell2018,catacora2019,lovell2020,catacora2021,catacora2023}. Here, `local' refers to optical potentials which are determined for specific reactions and beam energies.

% \begin{table*}[!htbp]
% \centering
% \begin{tabular}{|c|c|c|c|c|c|c|c|}
% \hline
%     Isotope & \makecell{$n_{tot}$\\(d$\sigma_{pp}$/d$\Omega$)} & \makecell{$E_{lab}$ (MeV)\\(d$\sigma_{pp}$/d$\Omega$)} & \makecell{$n_{tot}$\\($A_{y,pp}$)} & \makecell{$E_{lab}$ (MeV)\\($A_{y,pp}$)} & \makecell{$n_{tot}$\\(d$\sigma_{nn}$/d$\Omega$)} & \makecell{$E_{lab}$ (MeV)\\(d$\sigma_{nn}$/d$\Omega$)} & Sources \\
%     \hline
%     $^{90}$Zr & 195 & 14.71 - 49.35 & 57 & 16 - 20.25 & 96 & 10 - 24 & \cite{ball1964,varner1986,swiniarski1977,matsuda1967,vanderbijl1983,gray1966,mani1971,glashausser1969,bainum1978,wang1990} \\
%     \hline
%     $^{91}$Zr & 101 & 18.7 - 49.35 & 0 & N/A & 70 & 10 - 24 & \cite{ball1964,mani1971,blok1969,wang1990} \\
%     \hline
%     $^{92}$Zr & 125 & 19.4 - 49.35 & 26 & 20.25 & 70 & 10 - 24 & \cite{ball1964,swiniarski1977,mani1971,stautberg1966,glashausser1969,wang1990} \\
%     \hline
%     $^{94}$Zr & 109 & 19.4 - 49.35 & 0 & N/A & 68 & 10 - 24 & \cite{ball1964,mani1971,stautberg1966,wang1990} \\
%     \hline
%     $^{96}$Zr & 71 & 22.5 - 49.35 & 0 & N/A & 0 & N/A & \cite{ball1964,mani1971} \\
%     \hline
% \end{tabular}
% \caption{Summary of EXFOR data retrieved for calibration.}
% \label{tab:exfor_table}
% \end{table*}

More recently, uncertainty-quantified global optical potentials have been developed which are applicable across the entire nuclear landscape \cite{pruitt2023}, 
whereby the optical potential's mass, charge, and energy dependence are encoded in a large set of parameters, following the canonical parameterizations of Refs. \cite{koning2003, varner1991}. The Koning-Delaroche global potential (KD03) \cite{koning2003} was developed using experimental data across a wide range of energies and targets, and incorporating compound corrections at low incident energies. However, KD03 is not Lane-consistent \cite{lane1962isobaric}, i.e it does not consistently connect the proton and neutron potentials. 

%In contrast, Ref. \cite{bauge2001} presents a global optical potential developed with the desideratum of being Lane-consistent. However, due to considerations of dimensional simplicity and previous examples of uncertainty quantification studies, 

In contrast, the global Chapel Hill (CH89) \cite{varner1991} parameterization of the optical potential was developed over a narrower range of targets and beam energies, but is Lane-consistent, and thus utilizes a smaller set of phenomenologically-fitted parameters. In this study, we utilize the CH89 parameterization as a template for our Bayesian calibration. Ref. \cite{pruitt2023} introduces CHUQ, a global, uncertainty-quantified potential based on CH89 which informed our calibration procedure. It should be noted that due to  computational cost, so far no Bayesian uncertainty-quantified dispersive optical potential is available.

Global optical potentials do not describe all reactions with uniform accuracy, but are useful for describing average behaviors across the nuclear chart, and in particular reactions near the valley of stability. The present work explores the idea of a \textit{regional} potential, one that applies only to isotopes in the same mass and charge region (similar to the effective interactions of the shell model \cite{brown2006}). Such regional optical potentials would be developed with a specific application in mind, and should provide improved extrapolations away from the calibration isotopes to nearby targets. We specifically describe a Bayesian calibration using zirconium isotopic chain data, based on the potential form of CH89: the \textit{Chapel Hill Regional Potential} (CHiRP).

Zirconium nuclides are common fission products, for which there exists a relatively large corpus of experimental data. Following our calibration of the OMP parameters, we propagate the posteriors to compound nucleus reactions channels with {\sc talys} to make predictions of corresponding reaction cross sections, and compare these results to those obtained with CHUQ\footnote{The \textit{democratic} calibration of CHUQ was used throughout this work. For more details see Ref. \cite{pruitt2023}.}. This propagation was performed for neutron-induced reactions on $^{90}$Zr and $^{89}$Y.

In Sec. \ref{sec:calibration} we summarize the nuclear reaction theory used, namely the optical model and compound nucleus reactions. In Sec. \ref{sec:workflow} we explain the statistical methodology utilized. Sec. \ref{sec:data} presents the experimental data used to calibrate CHiRP. Sec. \ref{sec:results} presents our results, and conclusions are drawn in Sec. \ref{sec:conclusions}.

\begin{figure}[tbp!]
    \centering

    \begin{subfigure}{0.5\textwidth}
        \centering
        \includegraphics[width=\textwidth,keepaspectratio]{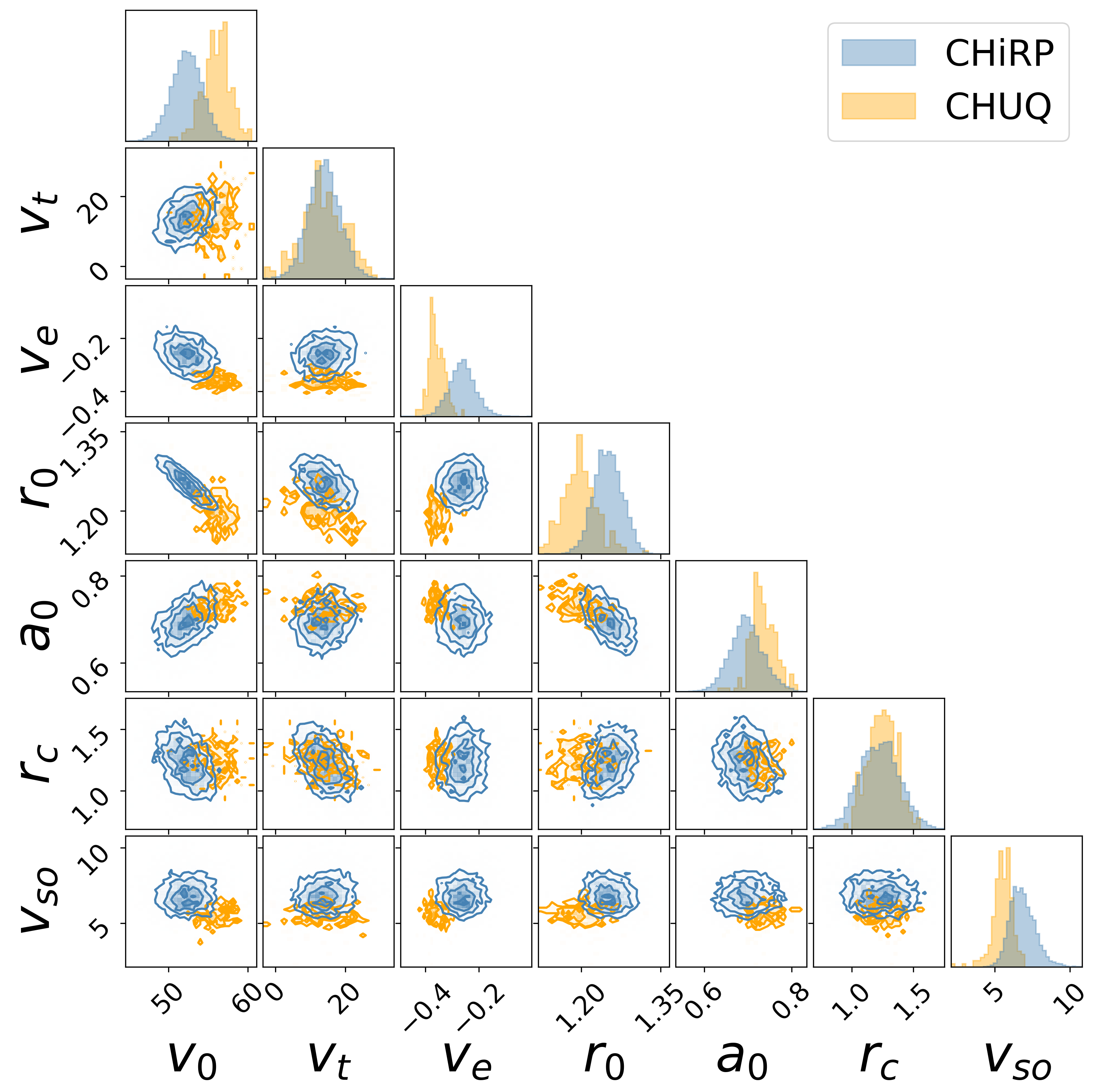}
        \caption{Real, central potential parameters and primary Coulomb radius parameter.}
        \label{fig:zr-real-corner}
    \end{subfigure}

    \vspace{0.5em}

    \begin{subfigure}{0.5\textwidth}
        \centering
        \includegraphics[width=\textwidth,keepaspectratio]{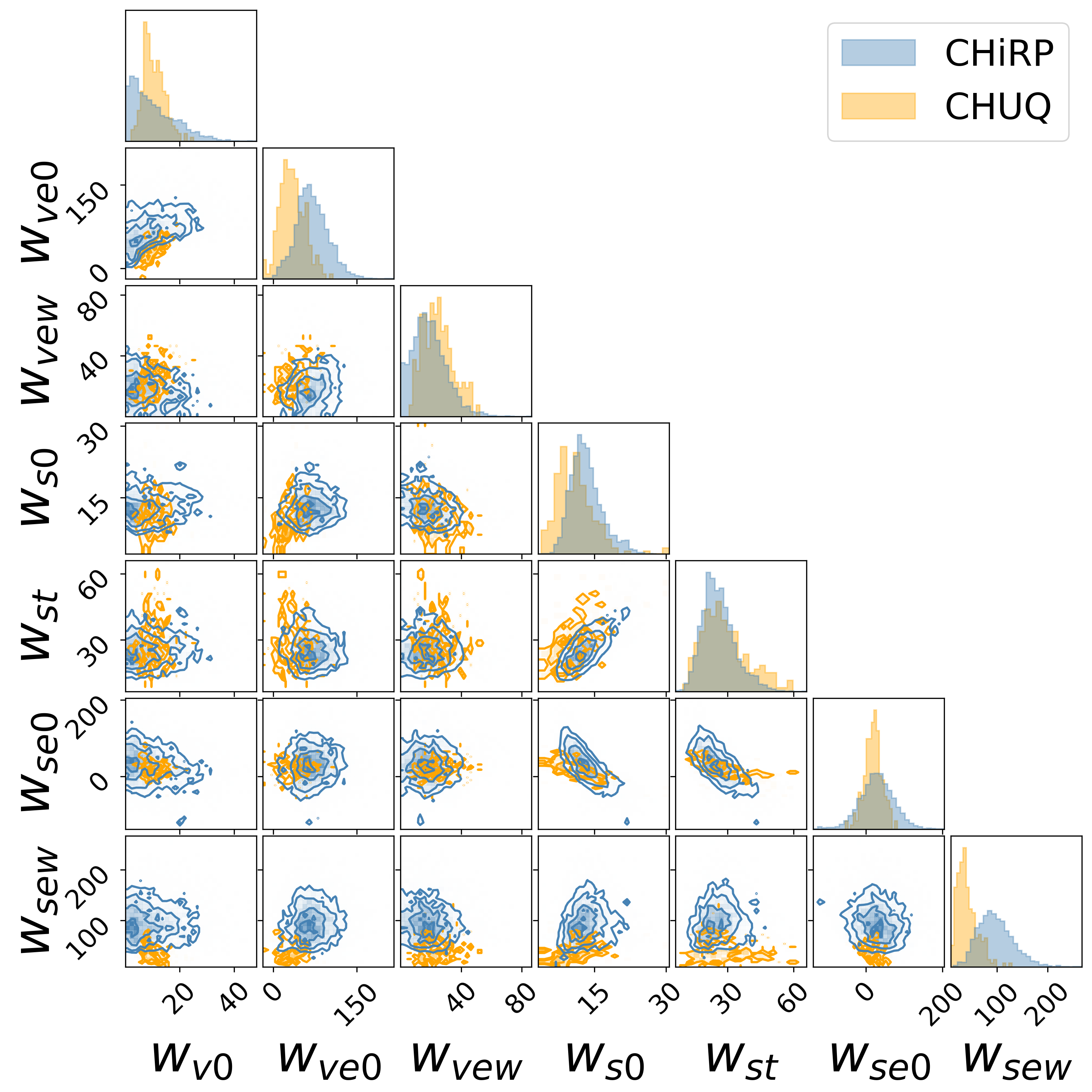}
        \caption{Imaginary potential parameters.}
        \label{fig:zr-imaginary-corner}
    \end{subfigure}

    \caption{Corner plots displaying marginal posterior distributions (along the diagonal) and their correlations (in the off-diagonal). CHiRP (blue) and CHUQ (orange) results are shown.}
    \label{fig:zr-corner-combined}
\end{figure}

\begin{table}
\centering
\begin{tabular}{|c|c|c|c|}
    \hline
    Parameter & CH89 & CHUQ & CHiRP \\
    \hline
    $v_0$ & $52.9$ & $56.19^{+1.43}_{-1.82}$ & $52.34^{+1.92}_{-1.94}$  \\
    %\hline
    $v_t$ & $13.1$ & $13.82^{+7.03}_{-5.25}$ & $13.98^{+4.5}_{-4.45}$ \\
    %\hline
    $v_e$ & $-0.299$ & $-0.36^{+0.03}_{-0.02}$ & $-0.26^{+0.05}_{-0.05}$ \\
    %\hline
    $r_0$ & $1.25$ & $1.20^{+0.03}_{-0.03}$ & $1.25^{+0.03}_{-0.03}$ \\
    %\hline
    $r_0^{(0)}$ & $-0.225$ & $-0.20^{+0.12}_{-0.13}$ & N/A\\
    %\hline
    $a_0$ & $0.69$ & $0.73^{+0.03}_{-0.02}$ & $0.7^{+0.04}_{-0.04}$ \\
    %\hline
    $w_{v0}$ & $7.8$ & $9.92^{+4.63}_{-2.92}$ & $8.35^{+10.89}_{-5.84}$ \\
    %\hline
    $w_{ve0}$ & $35$ & $33.15^{+25.03}_{-19.82}$ & $67.82^{+30.52}_{-24.81}$ \\
    %\hline
    $w_{vew}$ & $16$ & $24.00^{+11.32}_{-9.52}$ & $17.87^{+11.73}_{-10.1}$ \\
    %\hline
    $w_{s0}$ & $10$ & $10.59^{+3.99}_{-3.39}$ & $12.93^{+3.55}_{-2.76}$ \\
    %\hline
    $w_{st}$ & $18$ & $27.09^{+12.28}_{-8.72}$ & $25.16^{+7.95}_{-6.03}$ \\
    %\hline
    $w_{se0}$ & $36$ & $20.00^{+21.69}_{-20.82}$ & $28.95^{+42.95}_{-41.83}$ \\
    %\hline
    $w_{sew}$ & $37$ & $36.38^{+23.75}_{-13.66}$ & $93.88^{+36.59}_{-30.54}$ \\
    %\hline
    $r_w$ & $1.33$ & $1.32^{+0.08}_{-0.08}$ & $1.34^{+0.03}_{-0.03}$ \\
    %\hline
    $r_w^{(0)}$ & $-0.42$ & $-0.41^{+0.36}_{-0.32}$ & N/A\\
    %\hline
    $a_w$ & $0.69$ & $0.69^{+0.05}_{-0.05}$ & $0.64^{+0.06}_{-0.06}$ \\
    %\hline
    $v_{so}$ & $5.9$ & $5.58^{+0.52}_{-0.58}$ & $6.79^{+0.89}_{-0.73}$ \\
    %\hline
    $r_{so}$ & $1.34$ & $1.29^{+0.11}_{-0.11}$ & $1.25^{+0.1}_{-0.1}$ \\
    %\hline
    $r_{so}^{(0)}$ & $-1.2$ & $-1.12^{+0.45}_{-0.51}$ & N/A \\
    %\hline
    $a_{so}$ & $0.63$ & $0.61^{+0.04}_{-0.04}$ & $0.66^{+0.13}_{-0.09}$ \\
    %\hline
    $r_{c}$ & $1.24$ & $1.25^{+0.12}_{-0.12}$ & $1.23^{+0.16}_{-0.16}$ \\
    %\hline
    $r_{c}^{(0)}$ & $0.12$ & $0.13^{+0.09}_{-0.12}$ & N/A\\
    \hline 
    $\beta_{pp}$ & N/A & $0.31^{+0.10}_{-0.06}$ & $0.22^{+0.02}_{-0.02}$ \\
    $\beta_{nn}$ & N/A & $0.40^{+0.12}_{-0.10}$ & $0.21^{+0.02}_{-0.02}$ \\
    $\beta_{Ay,pp}$ & N/A & $0.77^{+0.11}_{-0.11}$ & $0.21^{+0.02}_{-0.02}$ \\
    \hline
\end{tabular}
\caption{OMP parameters for CH89, CHUQ, and CHiRP. Reported CHiRP values represent medians with the addition or subtraction of the difference between the median and the 16th and 84th percentile values, respectively. CHUQ values are reproduced from Ref. \cite{pruitt2023}. Potential well depth parameters have units of MeV, geometric parameters are in fm.}
\label{tab:parameters}
\end{table}

\begin{figure*}[!tbp]
    \centering
    \includegraphics[width=0.8\linewidth]{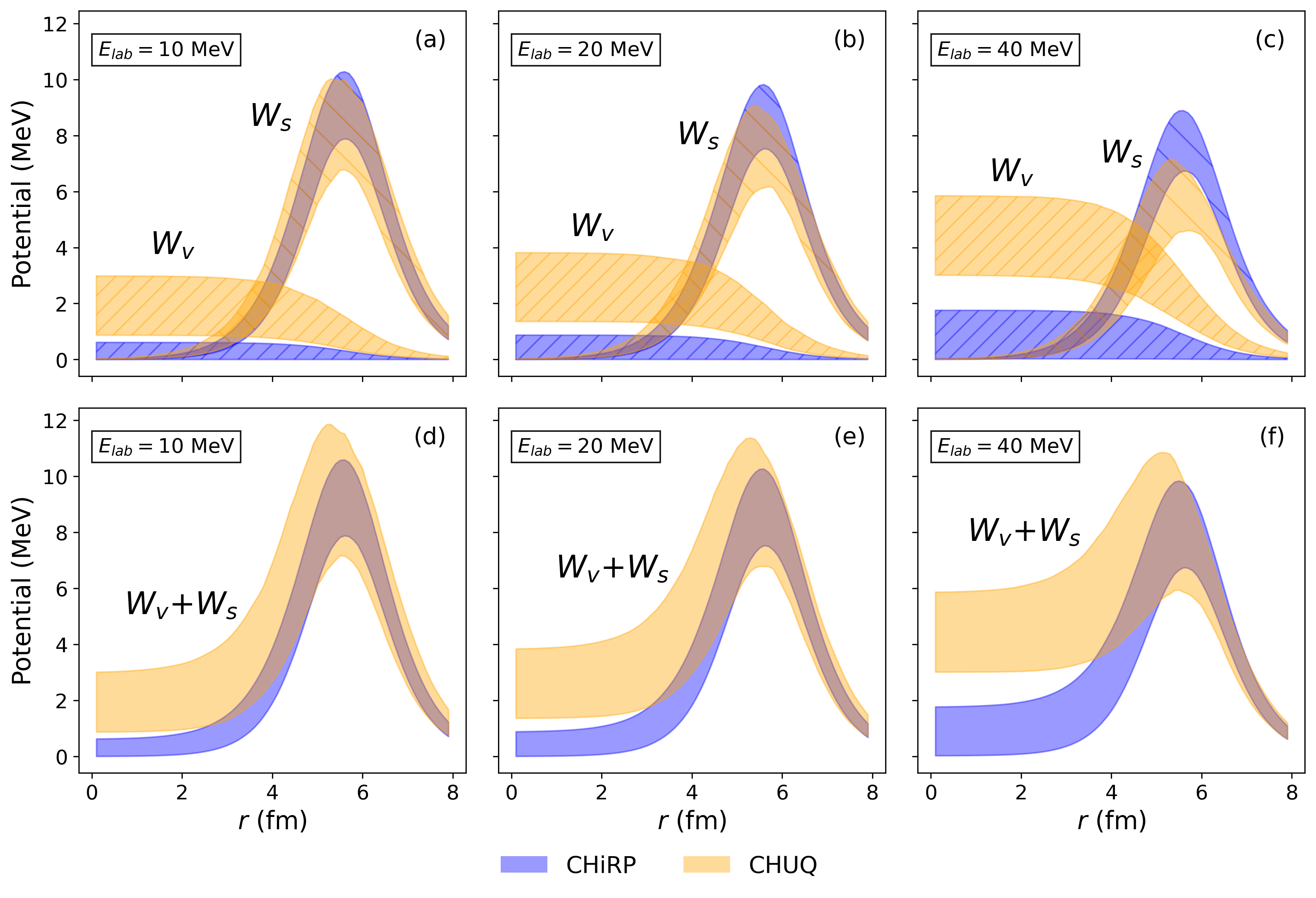}
    \caption{Radial behavior of the proton-$^{90}$Zr imaginary volume ($W_v$) and surface-peaked ($W_s$) potential terms (upper row) and their sums (lower row) for CHiRP (blue) and CHUQ (orange), at $E_{lab}$ = 10 (panels a,d), 20 (panels b,e), and 40 (panels c,f) MeV. Bands represent 1$\sigma$ confidence intervals.}
    \label{fig:imaginary_terms}
\end{figure*}

\begin{figure*}[!tbp]
    \centering
    \includegraphics[width=0.95\linewidth]{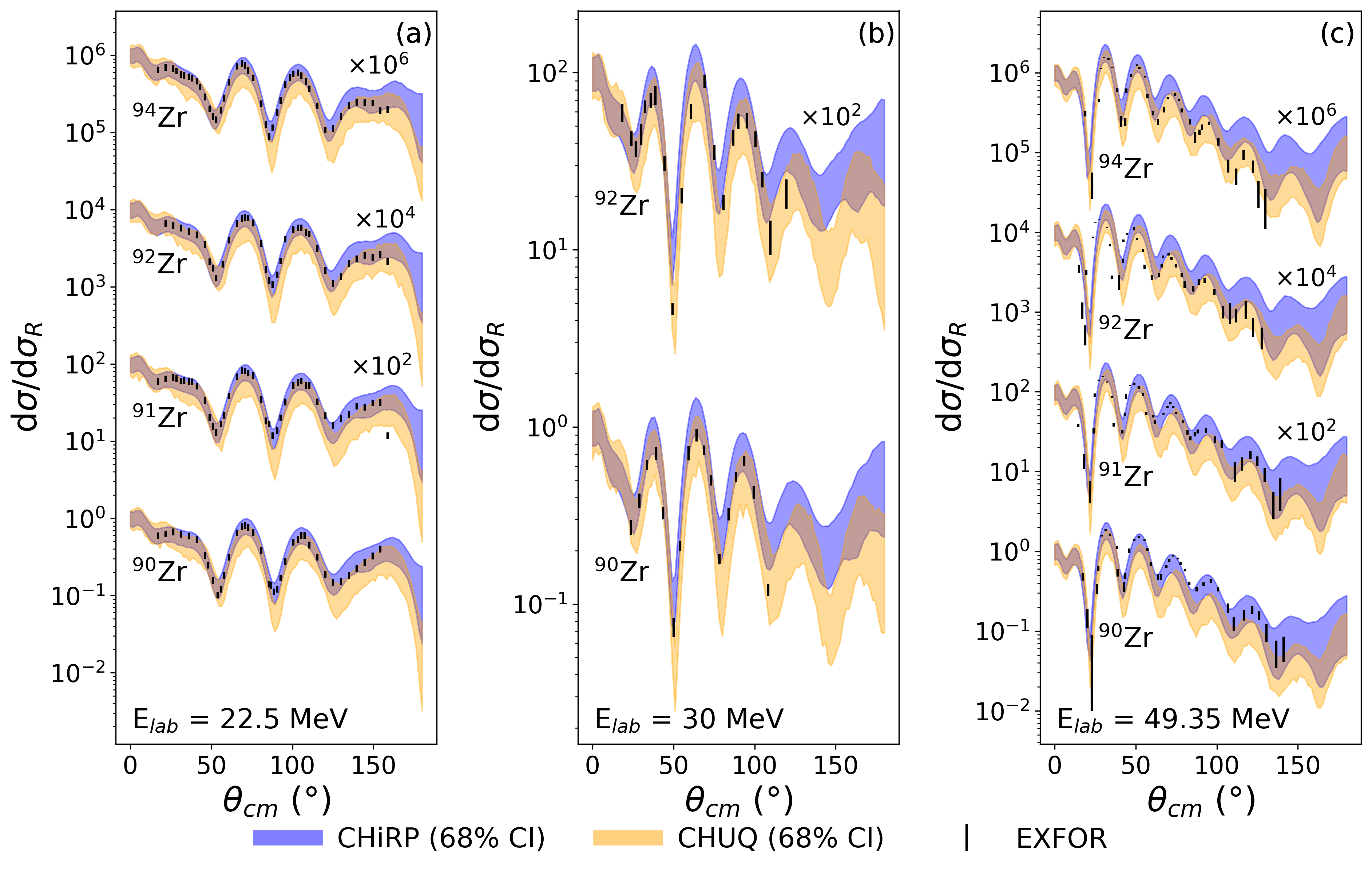}
    \caption{Proton elastic scattering angular distributions at (a) 22.5 MeV and (b) 30 MeV, and (c) 49.35 MeV incident proton energies for various zirconium targets. The confidence intervals (CIs) displayed represent the inner 68th percentile of predictions using CHiRP (blue) and CHUQ (orange). EXFOR experimental data are shown in black.}
    \label{fig:zr_p_scattering}
\end{figure*}

\begin{figure*}[!tbp]
    \centering
    \includegraphics[width=0.85\linewidth]{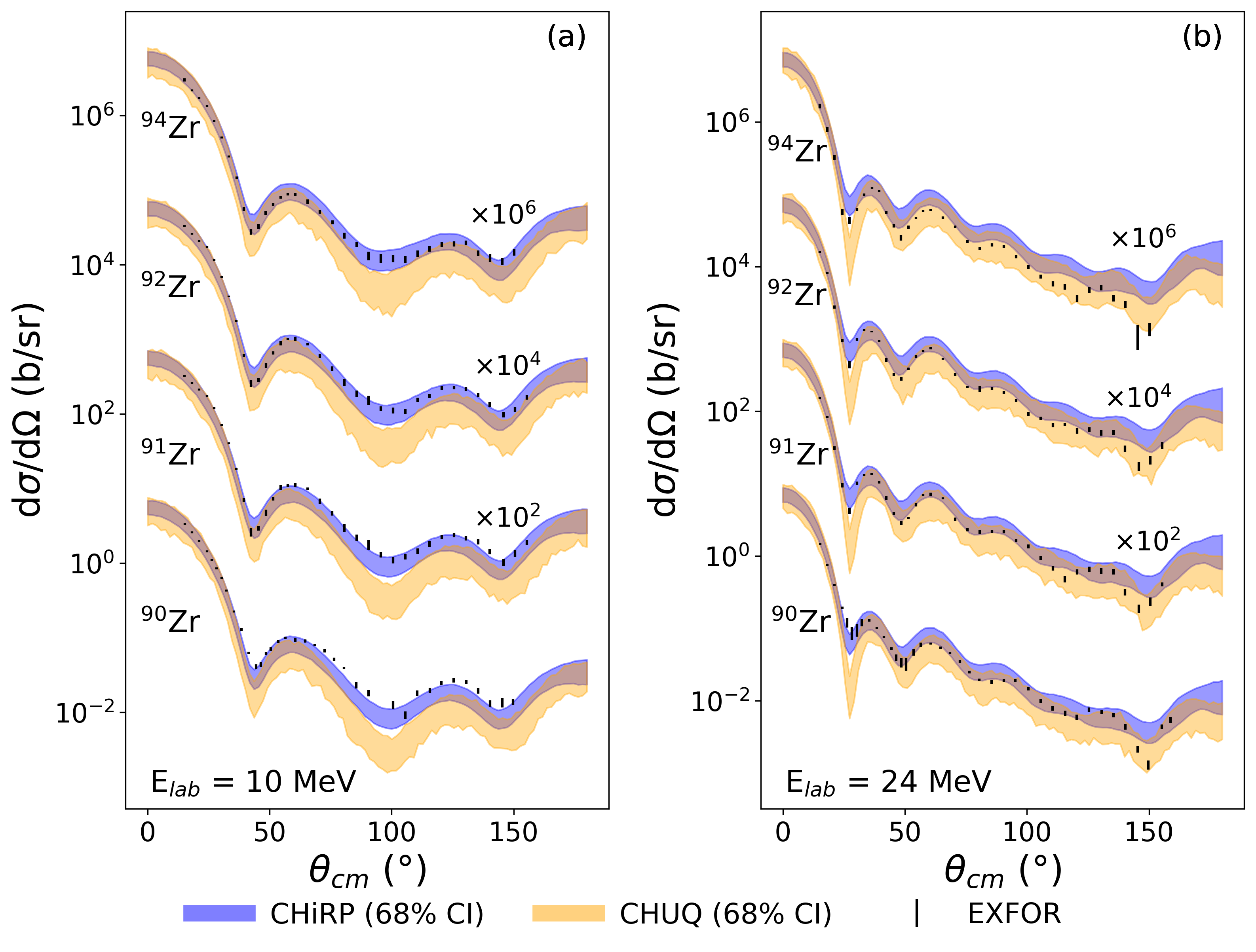}
    \caption{Neutron elastic scattering angular distributions at (a) 10 MeV and (b) 24 MeV incident neutron energies for various zirconium targets.}
    \label{fig:zr_n_scattering}
\end{figure*}

\begin{figure*}[!tbp]
    \centering
    \includegraphics[width=\linewidth]{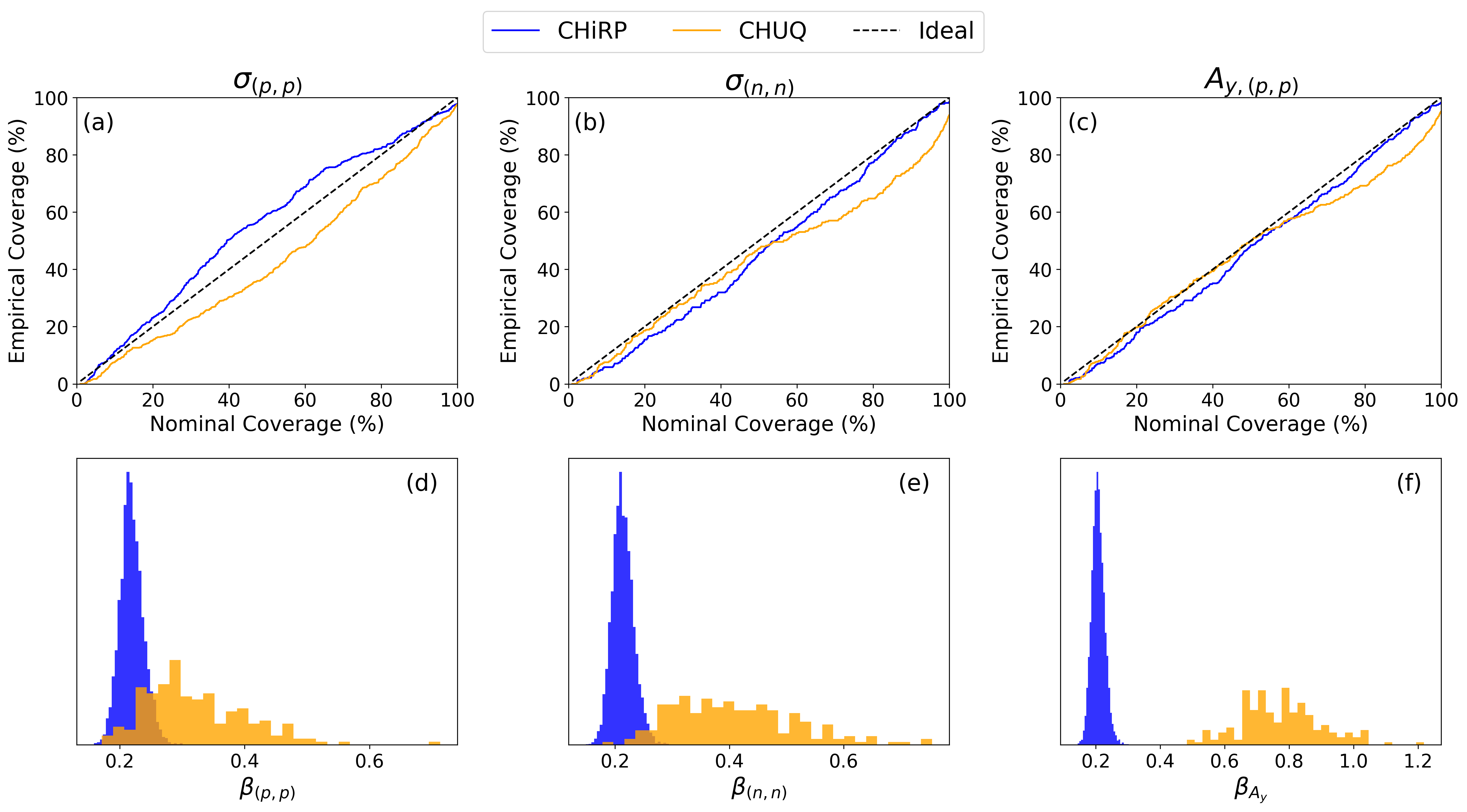}
    \caption{Panels (a), (b), and (c) show the average empirical coverage of CHiRP (blue) and CHUQ (orange) predictions across all proton differential cross sections, neutron differential cross sections, and proton analyzing power data. The black, dotted line indicates ideal coverage, where nominal coverage is equal to empirical coverage. Panels (d), (e), and (f) show the respective posterior distributions of the model error fraction parameters for CHiRP and CHUQ.}
    \label{fig:zr_ec}
\end{figure*}

\begin{figure}[!tbp]
    \centering
\includegraphics[width=0.5\textwidth,keepaspectratio]{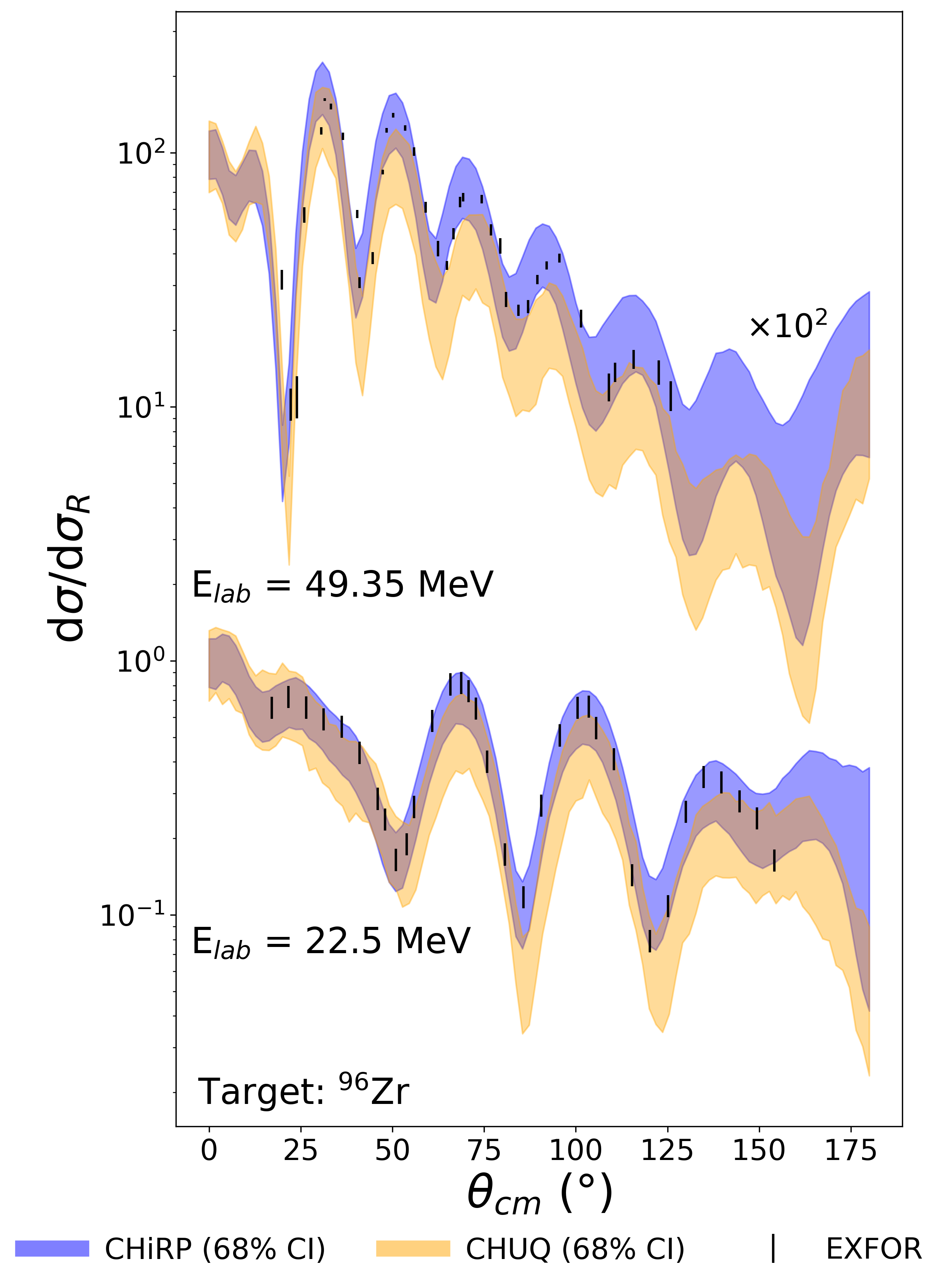}
    \caption{Proton elastic scattering angular distributions at 22.5 MeV and 49.35 MeV incident proton energies for a $^{96}$Zr target.}
    \label{fig:zr_ext}
\end{figure}

\begin{figure*}[!tbp]
    \centering
    \begin{subfigure}{0.49\textwidth}
        \centering
        \includegraphics[width=\textwidth,keepaspectratio]{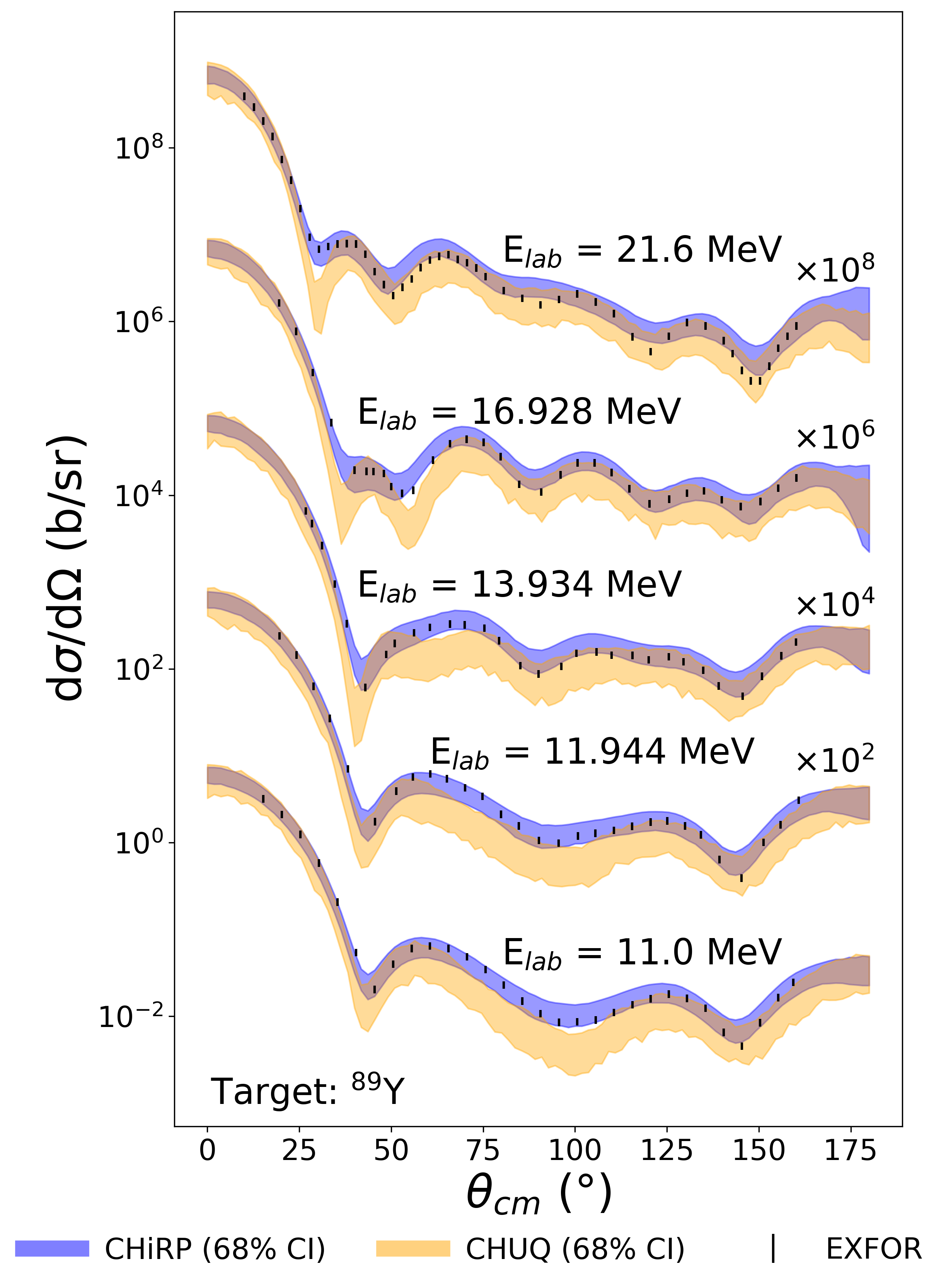}
        \caption{Neutron elastic scattering angular distributions at various energies for a $^{89}$Y target.}
        \label{fig:yttrium_scattering}
    \end{subfigure}
    \hfill
    \begin{subfigure}{0.49\textwidth}
        \centering
        \includegraphics[width=\textwidth,keepaspectratio]{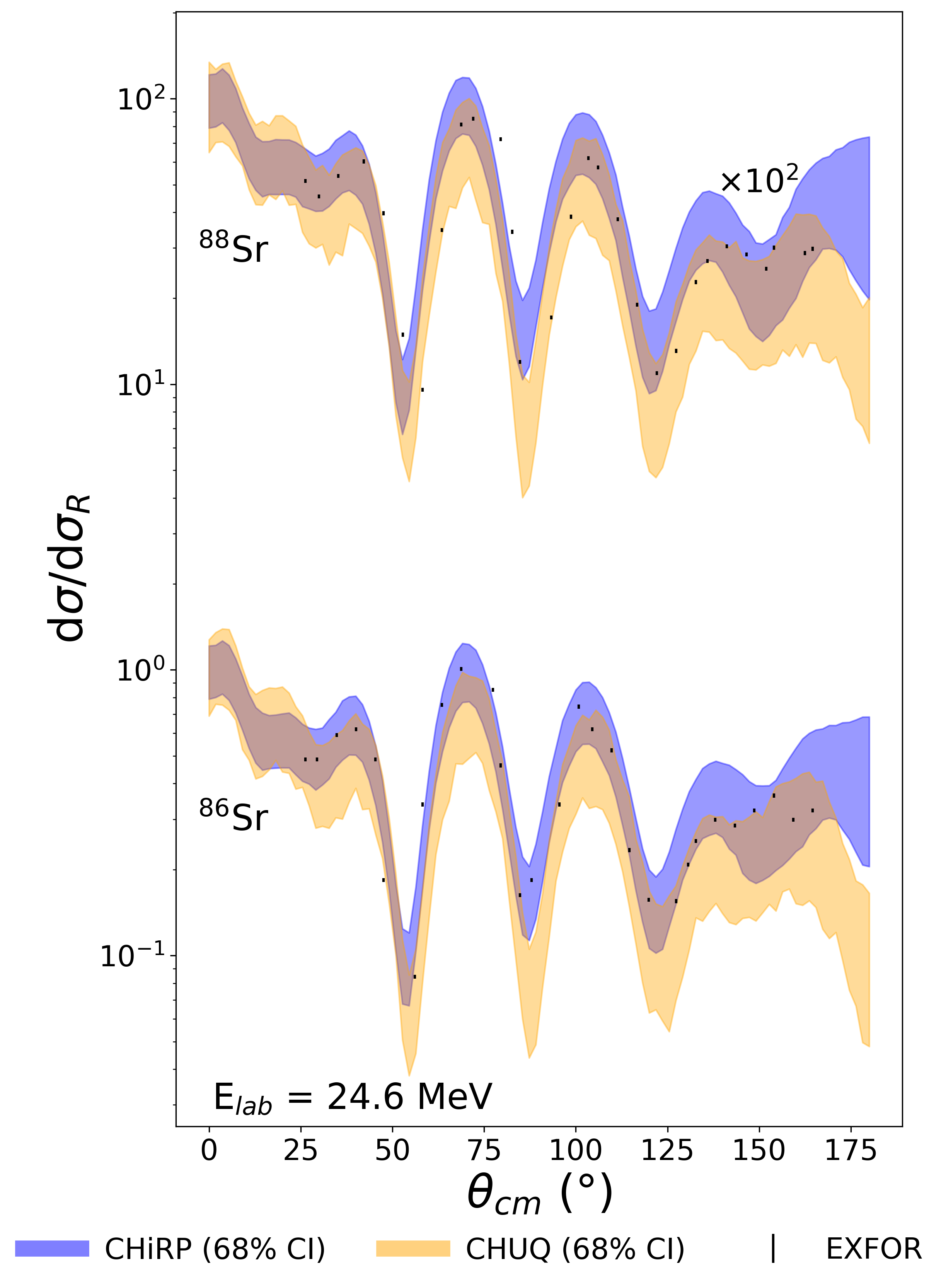}
        \caption{Proton elastic scattering angular distributions at 24.6 MeV for $^{86}$Sr and $^{88}$Sr targets.}
        \label{fig:sr_scattering}
    \end{subfigure}
\caption{Neutron and proton elastic scattering angular distributions for targets not included in the calibration data corpus.}
\end{figure*}

\begin{figure*}[!tbp]
    \centering
\includegraphics[width=0.85\textwidth,keepaspectratio]{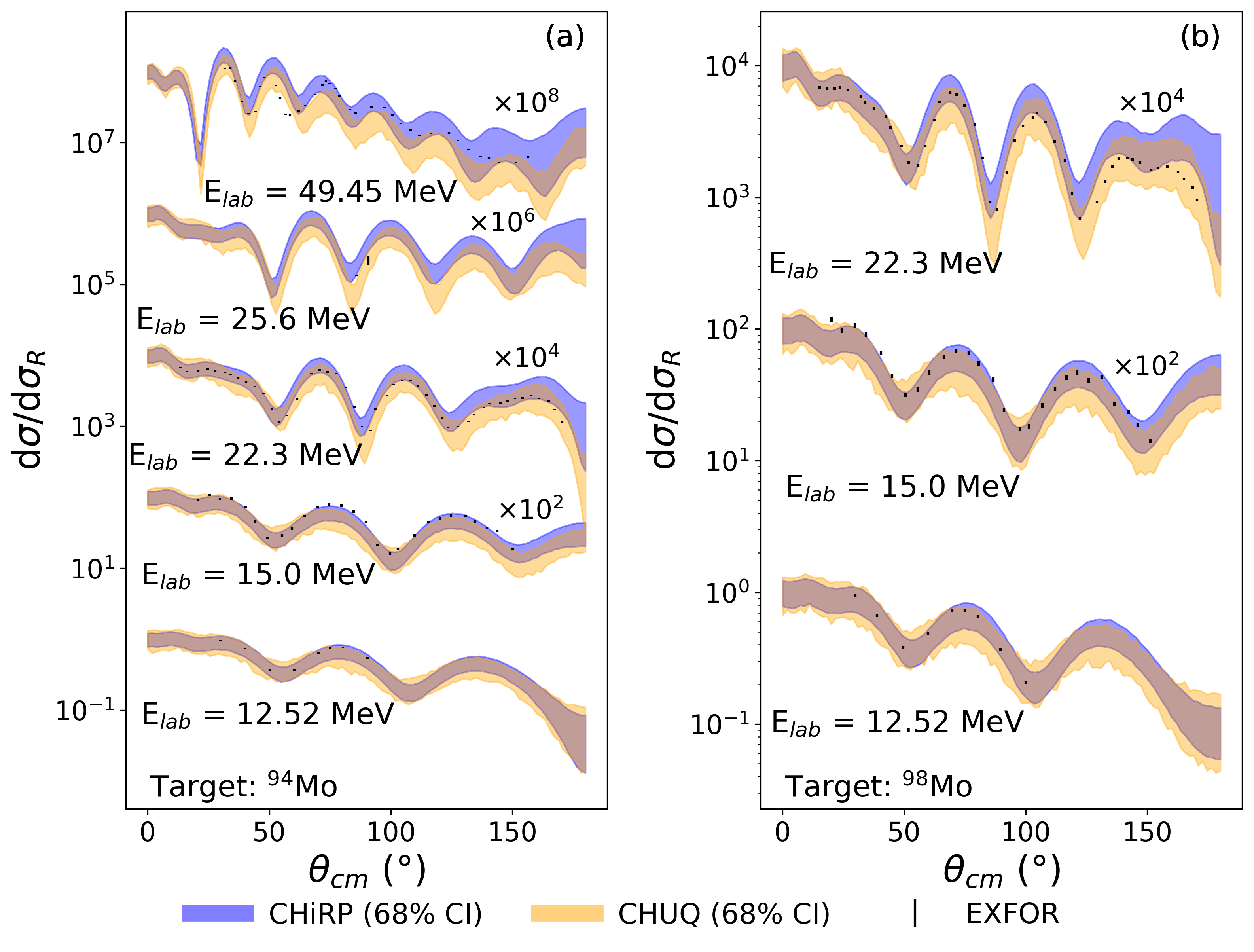}
    \caption{Proton elastic scattering angular distributions at various energies for (a) $^{94}$Mo and (b) $^{98}$Mo targets.}
    \label{fig:mo_scattering}
\end{figure*}

\begin{figure}[!htbp]
    \centering
\includegraphics[width=0.5\textwidth,keepaspectratio]{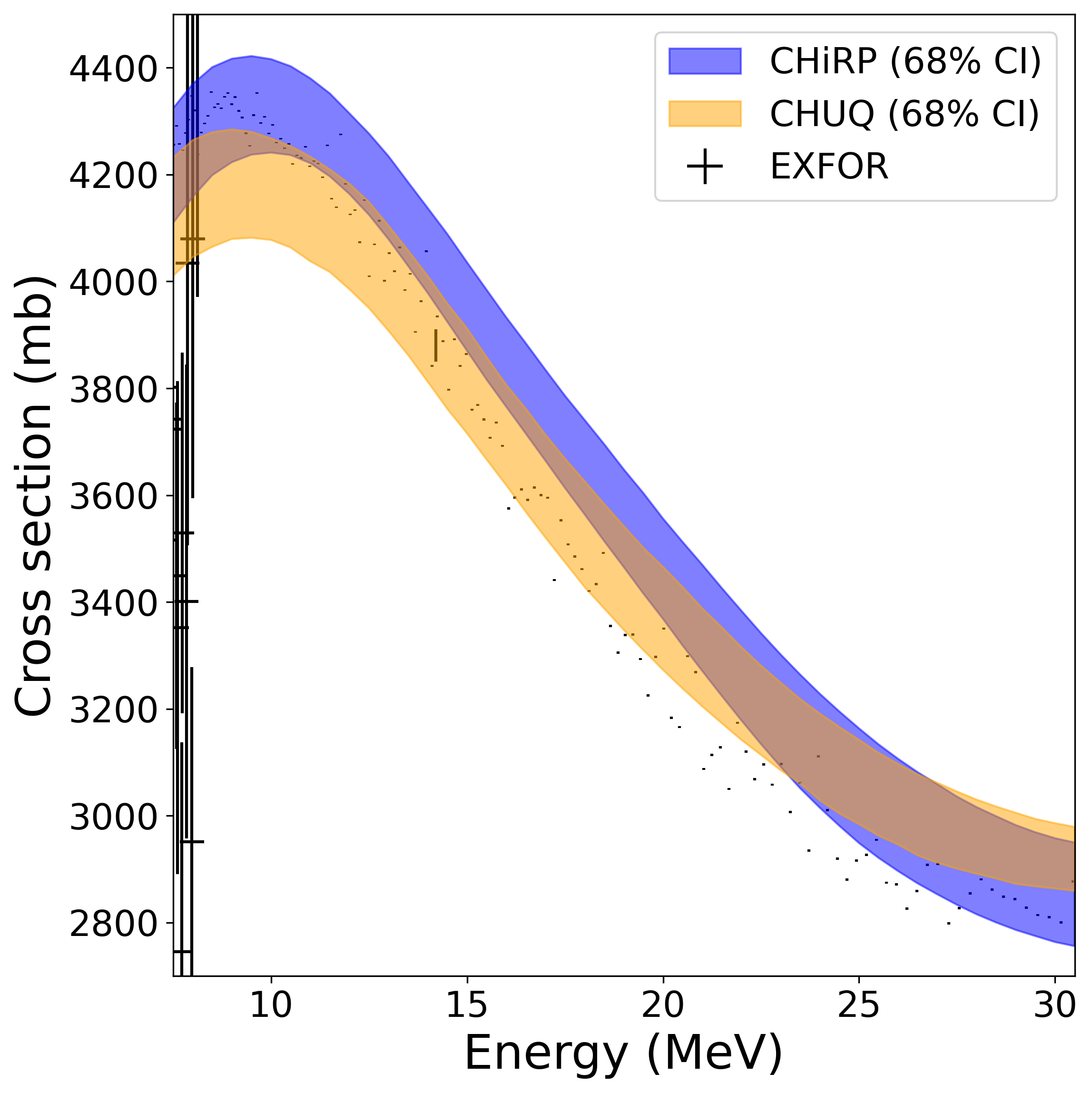}
    \caption{$^{90}$Zr$(n,tot)$ cross section results using CHiRP (blue) and CHUQ (orange).}
    \label{fig:zr_total}
\end{figure}

\begin{figure*}[!tbp]
    \includegraphics[width=\textwidth,keepaspectratio]
    {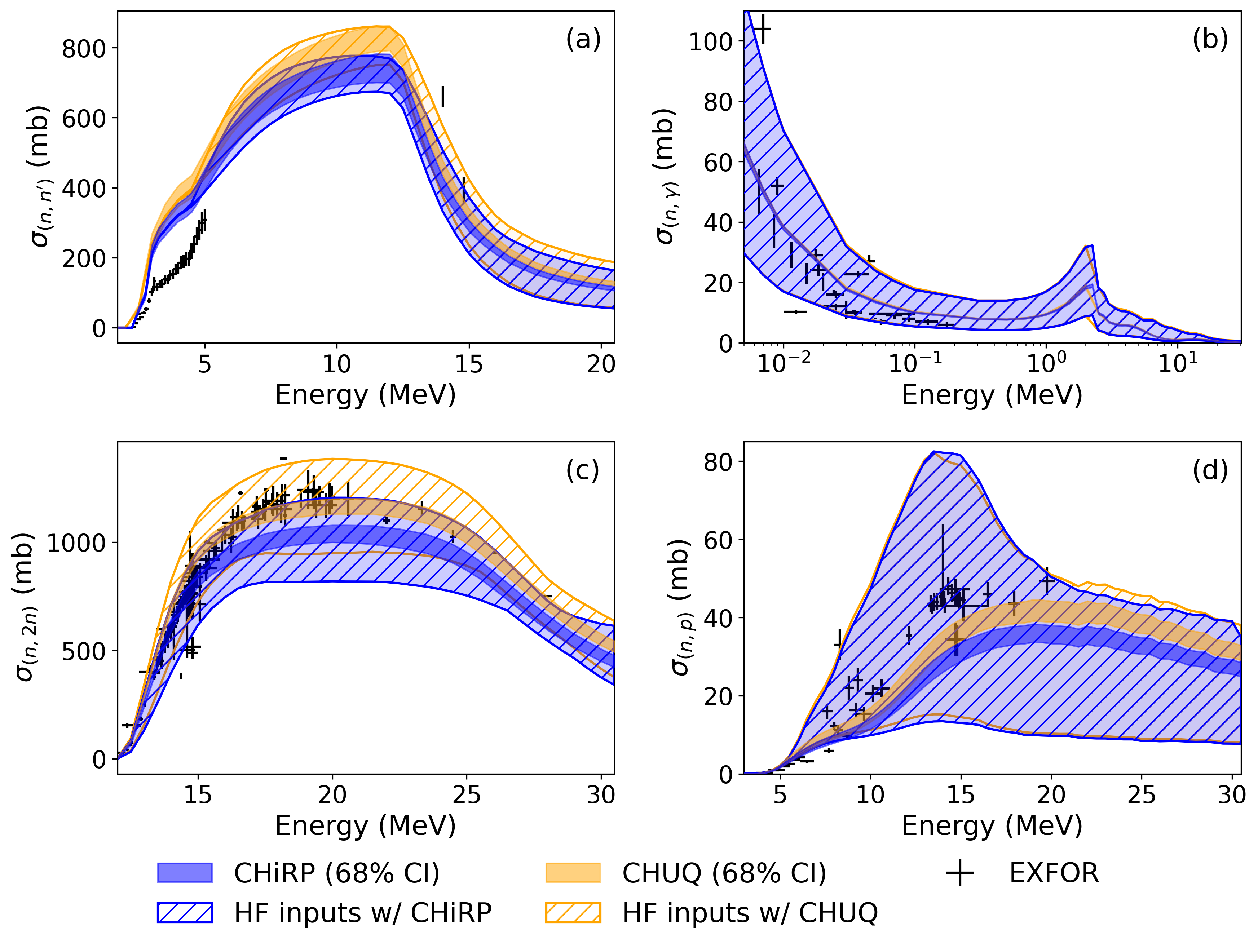}
    \caption{(a) $^{90}$Zr$(n,n')^{90m}$Zr, (b) $^{90}$Zr$(n,\gamma)^{91}$Zr, (c) $^{90}$Zr$(n,2n)$$^{89}$Zr, and (d) $^{90}$Zr$(n,p)$$^{90}$Y {\sc talys} cross section calculations. CHiRP and CHUQ 68\% CI shown in solid blue and orange, respectively. Light blue and hatched light orange respectively show results of varying all nuclear level density, gamma strength function, and pre-equilibrium models in {\sc talys} with central CHiRP and CHUQ OMP inputs. EXFOR data are shown in black.}
    \label{fig:zr_combined}
\end{figure*}

\begin{figure}[!tbp]
    \centering
\includegraphics[width=0.5\textwidth,keepaspectratio]{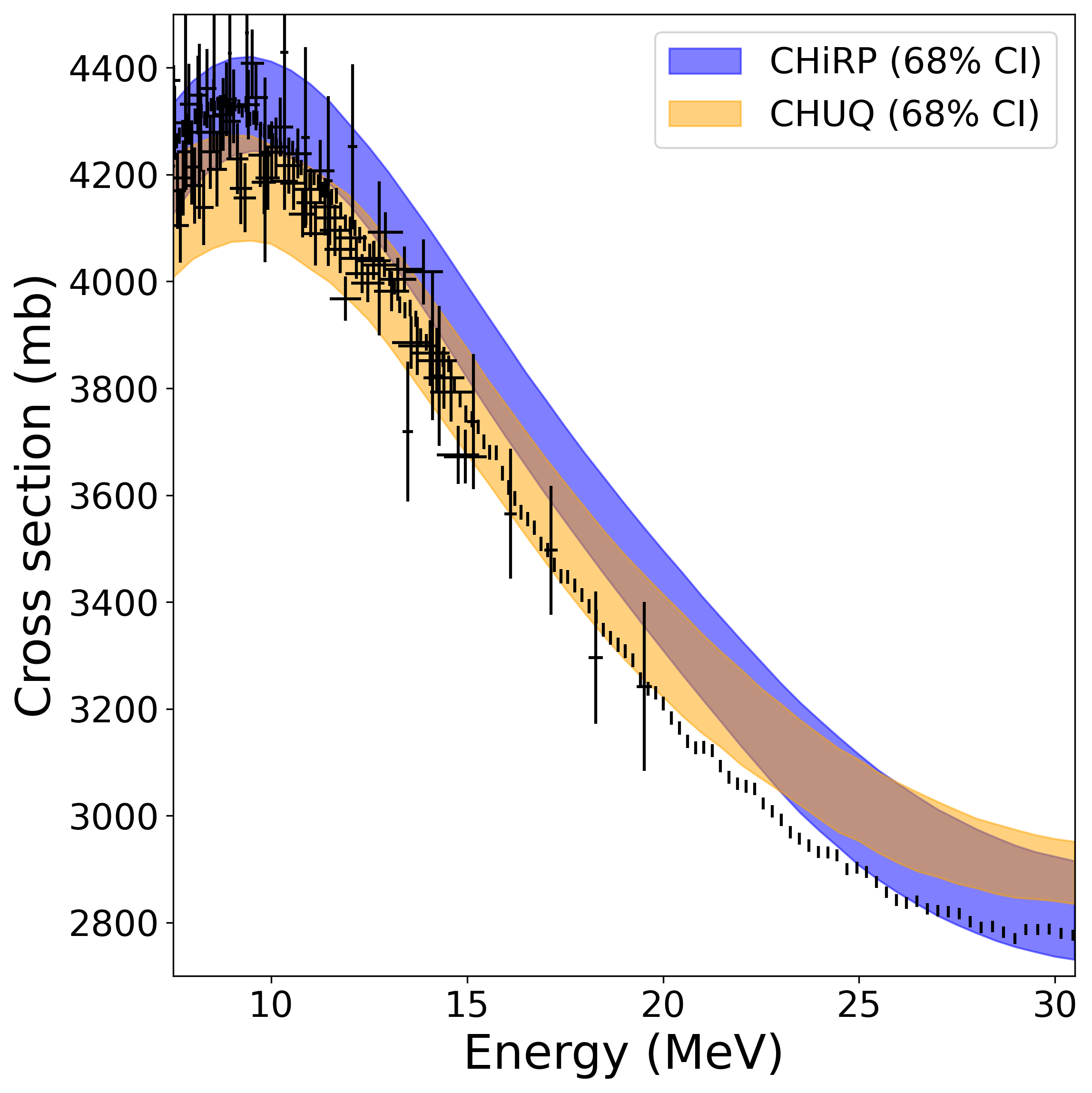}
    \caption{$^{89}$Y$(n,tot)$ cross section results using CHiRP (blue) and CHUQ (orange).}
    \label{fig:y_total}
\end{figure}

\begin{figure*}[!tbp]
    \includegraphics[width=\textwidth,keepaspectratio]
    {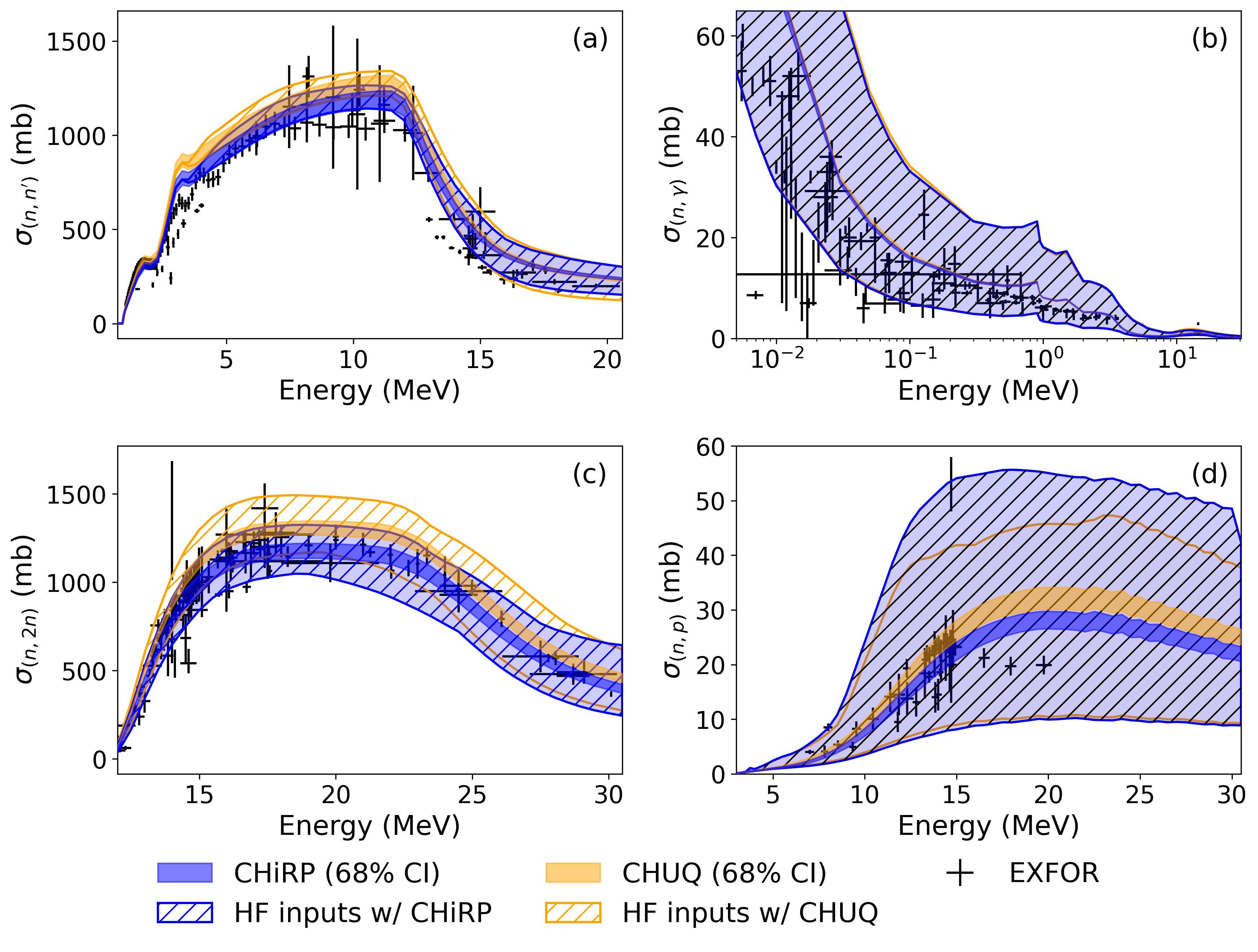}
    \caption{(a) $^{89}$Y$(n,n')^{89m}$Y, (b) $^{89}$Y$(n,\gamma)^{90}$Y, (c) $^{89}$Y$(n,2n)^{88}$Y, and (d) $^{89}$Y$(n,p)^{89}$Sr {\sc talys} cross section calculations.}
    \label{fig:y_combined}
\end{figure*}

\section{\label{sec:calibration}Theory}
\subsection{\label{sec:CHiRP}Optical Potentials}

The nucleon-nucleus OMP is an effective two-body nuclear potential that characterizes the interactions between a projectile nucleon and a target nucleus \cite{hodgson_1971}. Typically, OMPs are non-Hermitian, where the real component captures the diffraction in elastic scattering, and the imaginary (or \textit{absorptive}) component represents those transitions that remove flux from the elastic channel. Here, we consider a central OMP valid for spherical nuclei, following CH89 \cite{varner1991},

\begin{eqnarray}
U(r, E)=&&\mathcal{V}_{r}-i(\mathcal{W}_v+\mathcal{W}_s)-\mathcal{V}_{so}(\ell\cdot\sigma)+\mathcal{V}_C,
\label{CH89OMP}
\end{eqnarray}
where $\mathcal{V}_{r}$ and $i\mathcal{W}_v$ are respectively real and imaginary volume terms, $\mathcal{W}_s$ is a surface-peaked absorptive term, and $\mathcal{V}_{so}$ is the real spin-orbit term. In addition, protons experience the Coulomb force according to $\mathcal{V}_C$. 

CH89 uses phenomenologically-fitted parameters in order to reproduce experimental observations. For example, take the real, volume term of the CH89 potential,

\begin{eqnarray}
    \mathcal{V}_r(r,E) 
    &&= -V_r(E,v_0,v_e,v_t)f(r,R_0,a_0)
    \nonumber\\
    &&= -(v_0+v_e\Delta E\pm a_{sym} v_t)f(r,R_0,a_0).
    \label{real_central_eq}
\end{eqnarray}

\noindent In this equation, we have a depth function, $V_r(E,v_0,v_e,v_t)$, which determines the scale of the nuclear potential as a function of lab-frame projectile energy, \textit{E}, with three fitted parameters, $v_0$, $v_e$, and $v_t$. $f(r,R_0,a_0)$ is the Woods-Saxon form factor, which dictates how the potential varies with the center-of-mass projectile-target separation, $r$. It is parameterized by the radius $R_0$ and diffuseness $a_0$. $\Delta E$ is the Coulomb-shifted energy, corresponding to the difference between the scattering energy and the Coulomb energy ($E_C = 6Ze^2/5R_C$ for protons and $E_C=0$ for neutrons, with $Z$ being the atomic number of the target nucleus, $e$ the elementary charge, and $R_C$ the target's Coulomb radius). $a_{sym}$ is the nucleon asymmetry of the target, defined as $(N-Z)/A$, and the upper (lower) sign in Eq. (\ref{real_central_eq}) corresponds to the case of incident protons (neutrons). This form makes CH89 Lane-consistent, due to its isovector dependence. $v_t$ governs the asymmetry dependence of $\mathcal{V_r}$, responsible for the extrapolation to neutron-rich isotopes. 

CHiRP consists of 18 calibrated parameters. This is three fewer than CH89, as we simplify the expression for calculating radii. A complete description of CHiRP is provided in Appendix \ref{app:chirp}. 

When calibrating CHiRP, optical model calculations were performed with the just-in-time compiled $\mathcal{R}$-matrix solver \texttt{jitR} \cite{jitr}. This software package determines scattering wavefunctions using an $\mathcal{R}$-matrix approach, solving the Bloch-Schr\"odinger equation on a Lagrange-Legendre mesh. Just-in-time compilation is achieved through \texttt{numba} \cite{numba}, which adapts Python functions to optimized machine code. \texttt{jitR} further optimizes its computations by defining and pre-computing variables of the reaction calculation. Tests involving the CHUQ parameter set indicated that \texttt{jitR}'s default solver settings (matching radius $R_{match}=14$ fm, a maximum of 50 partial waves, and basis number $N_{basis}=20$) lead to converged results for the reactions under consideration. \texttt{jitR} does not perform compound nucleus calculations, but produces elastic scattering differential cross sections, total reaction differential cross sections, and analyzing powers.

\subsection{\label{sec:cn}Compound Nucleus Reactions}

Compound nucleus reactions are those in which a nucleon is captured by a nucleus, equilibrating its energy among the interior nucleons, resulting in an excited \textit{compound} state of the composite $A+1$ system. This combined system may then de-excite through any number of energetically available exit channels. These include emission of a photon, nucleon, or other composite particle, and in some circumstances, fission.

Compound processes proceed along longer timescales than direct elastic reactions. According to Bohr's independence hypothesis \cite{bohr1936}, due to the statistical nature by which energy is distributed throughout the nucleus during these reactions, the nucleus `forgets' the identities of the entrance channel constituents, and so the choice of decay channel is independent from the entrance channel, aside from considerations of incident energy, spin, and parity. 

Due to the statistical nature of the compound nucleus, Hauser-Feshbach (HF) models are often adopted for describing these reactions. This model factorizes the cross section into independent contributions from the entrance channel and all possible exit channels. We may write the cross section of the decay of a compound nucleus from entrance channel $a$ via exit channel $a'$ as:

\begin{equation}
   \sigma_{a'} = \sum_{a} \sigma_{a}\mathcal{B}_{a':a}
\end{equation}

\noindent where $\sigma_a$ is the compound nucleus formation cross section via entrance channel $a$, and the branching ratio from initial state $a$ to $a'$ is:

\begin{equation}
\label{eq:branching_ratio}
    \mathcal{B}_{a':a} = \sum_{L,J}\frac{T_{L,J,a':a}}{\sum_{L'',J'',a''}T_{L'',J'',a'':a}}
\end{equation}

$T$ represents transmission coefficients between initial state $a$ and the given final state. Coefficients are determined for each partial wave, $(L,J)$. These coefficients are uniquely determined by the optical potential through the scattering matrix, $\mathbf{S}_{L,J,aa}$. The relation between these is:

\begin{equation}
    T_{L,J,a':a} = 1 - \left | \mathbf{S}_{L,J,aa} \right |^2
\end{equation}

\noindent Through this relation, optical model uncertainty can be propagated through the compound nucleus reaction process. 

The denominator of Eq. (\ref{eq:branching_ratio}) sums over all allowed decay channels of the compound nucleus. In practice, this requires the inclusion of further considerations, such as level density models and photon strength functions \cite{nunes2009}.

In this study, the reaction code {\sc talys} (version 2.0) was used to propagate optical model uncertainty and perform compound nucleus calculations. {\sc talys} is capable of simulating a vast array of nuclear reactions and processes over a large range of incident energies, masses, and projectiles. This includes direct reactions, pre-equilibrium reactions, and compound nucleus decay.

%\IB{In addition to propagating optical uncertainty, we endeavored to provide a crude estimate of the scale of uncertainties due to HF model inputs aside from the optical potential. To achieve this, we obtained results using each combination of nuclear level density, gamma-ray strength function, and pre-equilibrium model included in {\sc talys}, using CHiRP's central values. This same procedure was repeated using the CHUQ potential, for comparison purposes. The models used to obtain this data, as they are listed in the {\sc talys} documentation, are shared in Appendix \ref{app:hf_inputs}.}

%Combining the results of the calibration procedure with {\sc talys} calculations required converting our optical model parameters to a form accepted by {\sc talys}. As {\sc talys} expects a KD03 form for the optical potential, which has a single radius parameter, $r_{KD,x}$, the two parameters used to define the radius terms in CH89 ($r_x$ and $r_x^{(0)}$) were converted to the appropriate form according to $r_{KD,x}=(r_xA^{1/3}+r_x^{(0)})/A^{1/3}$, where $A$ is the target mass number. In addition, the real surface-peaked and imaginary spin-orbit potential depths defined in {\sc talys} were set to 0, consistent with our calibration and CH89.

\section{\label{sec:workflow}Statistical Methodology}

We consider a physical model that makes predictions for specific observables $y_m(x_i;\vars)$  which depend on physical variables $x_i$ and parameters $\vars$. The goal is to infer from observations $\mathcal{D} \equiv \{x_i,y_i\}_{i=1}^N$ the parameters $\vars$ such that the model $y_m(x;\vars)$ is reliably predictive for new $x$. Following \cite{kennedy2001bayesian}: $y_i + \epsilon_i = y_m(x_i;\vars) + \delta(x_i) $ which include the data uncertainties $\epsilon_i$, and the model discrepancy $\delta(x_i)$.
For our application to the nucleon-nucleus optical model, $y_i$ corresponds to either the differential cross section ($d\sigma/d\Omega$) or analyzing power ($A_y$) at a given bombarding energy $E$, target ($A,Z$), and projectile ($n$ or $p$), while $x_i$ corresponds to the scattering angle $\theta$, as well as bombarding energy, $E$, target isotope, reaction, and observable (differential cross section or analyzing power). We are especially interested in calibrating $\alpha$ to a range of target isotopes along a single isotopic chain $Z$, and to reliably extrapolate along $(N-Z)/A$ to neighboring, neutron-rich targets. 

We assume the experimental uncertainty is distributed according to a multivariate normal, and that the covariance is diagonal, $\Sigma_{ij} = \delta_{ij} \sigma^2_i$, with $\sigma_i$ being the reported statistical uncertainty of $y_i$. Given that they are not always reported, we ignore systematic errors in the measurements.

%Following the approach in the previous Bayesian global optical model calibration \cite{pruitt2023}, 
We assume the model discrepancy is normally distributed and expressed as:
\begin{equation}
    \Sigma^{\text{model}}_{ij} = \delta_{ij} \left( \beta y_m(x_i;\vars) \right)^2,
\end{equation}
which sets the standard deviation of the model discrepancy at $x_i$ to be some fraction $\beta$ of the model prediction $y_m$ (note that this differs from CHUQ which uses the average of $y_i$ and $y_m$).  In our study, the hyperparameter $\beta$ is  determined through calibration.
This simplified model discrepancy neglects correlations.

In practice, we split up our dataset $\mathcal{D}$ by observable class $l$ (according to projectile $n$ or $p$, and whether $y$ describes differential cross sections or analyzing powers), and by projectile. Note that we introduce a separate $\beta_l$ hyperparameter for each subset.
Given these definitions, the likelihood, $\mathcal{L}$, that the model, for a given set of $\vars$, describes one data subset $\mathcal{D}_l$, also follows a multivariate normal with diagonal covariance:

\begin{align}
\begin{split}
    \mathcal{L}(\mathcal{D}_l | \vars, \beta_l) &= \frac{1}{( (2 \pi)^{N_l} |\mathbf{V}_l| )^{1/2}} \exp{(-\frac{1}{2} \mathbf{r}^\top \mathbf{V}_l^{-1} \mathbf{r} )} \\
    r_i &\equiv y_i - y_m(x_i;\vars) \\
    V_{l,ij} &\equiv \Sigma_{ij} + \Sigma_{l,ij}^{\text{model}} =  \delta_{ij} \left( \sigma_i^2 + \left( \beta_l y_m \right)^2\right),
\end{split}
\end{align}

\noindent
and the total likelihood for all the independent datasets combine trivially: 
$\mathcal{L}(\mathcal{D} | \vars, \boldsymbol{\beta}) = \prod_l \mathcal{L}(\mathcal{D}_l | \vars, \beta_l)$.

Given priors $p(\vars)$ and $p(\boldsymbol{\beta}) \equiv \prod_l p(\beta_l)$, Bayes' theorem relates this total likelihood to the posterior distribution:
\begin{equation}
    p(\vars,\boldsymbol{\beta} | \mathcal{D} ) p(\mathcal{D}) = \left[ \prod_l \mathcal{L}(\mathcal{D}_l | \vars, \beta_l) \right] p(\bm{\beta}_l)p(\vars) .
\end{equation}
Note that the priors are truncated at patently unphysical values, e.g. a lower limit of 0 for radius parameters.

\noindent
However, in this work, we use a modified form of Bayes' theorem, inferring, instead of the posterior $p(\vars,\boldsymbol{\beta} | \mathcal{D} )$, the power-posterior, $p_\lambda(\vars,\boldsymbol{\beta} | \mathcal{D} )$, defined as \cite{friel2008marginal,grunwald2017inconsistency}:
\begin{equation}
    p_\lambda(\vars,\boldsymbol{\beta} | \mathcal{D} ) p(\mathcal{D}) =  \left[ \prod_l \mathcal{L}(\mathcal{D}_l | \vars, \beta_l) \right]^\lambda p(\vars) p(\boldsymbol{\beta}),
\end{equation}
\noindent
% \begin{equation}
%     p(\vars | \mathcal{D} )  =   \frac{\mathcal{L}(\mathcal{D} | \vars)  p(\vars)}{p(\mathcal{D})}
% \end{equation}
with $\lambda < 1$. This is equivalent to re-scaling the log-likelihood such that the weight of the likelihood is reduced relative to the prior. After a brief study (see Appendix \ref{app:lambda}) we chose to be consistent with \cite{pruitt2023} and used $\lambda = k/N$, where $k$ is the number of parameters in the model (the size of $\vars$), and $N$ is the total number of data points. We find that this scaling regularizes the sampling-inversion problem. 

With the amount of data used to calibrate CHiRP, we arrived at a scaling factor of $\lambda\approx0.02$ (we additionally trialed other values of $\lambda$, and these results are presented in Appendix \ref{app:lambda}).

A nested sampling \cite{skilling2006} procedure was adopted to construct our posterior distributions. These can then be used to obtain the posterior predictive distributions for the physics observables of interest, e.g. cross sections. In this work, we used \texttt{RxMC} \cite{rxmc} to perform the calibration. This is a modular Python library that allows a user to perform Bayesian calibrations of an arbitrary physical model, and is particularly suited to calibrations of nuclear reaction models due to its integration with \texttt{jitR}.

Model discrepancy error is propagated through the posterior predictive distributions by adding Gaussian noise proportional to the calibrated model error fractions. A prediction, $y_{l,m}$, is first made using a set of calibrated OMP parameters, then this prediction is used as the mean of a Gaussian distribution with standard deviation equal to the absolute product of the prediction and corresponding error fraction, $|\beta_{l} y_m|$. This procedure was followed for all predictions of observables included in the calibration corpus, however model discrepancy is left unaccounted for when calculating other observables, e.g. total reaction cross sections. 

Independent normal distributions were used as parameter prior distributions, centered at the original global CH89 parameter values. The standard deviations for these distributions ranged from 25\% to 150\% of the corresponding central value. This choice allows a large range of parameter values to be explored. The details of the prior distributions are presented in Appendix \ref{app:nested_sampling}.

Empirical coverage, which tells us the percentage of experimental data contained by the nominal coverage interval \cite{gelman2013}, is an important test to gauge the reliability of the uncertainties obtained. Ideally, nominal coverage of the calculation should agree with the empirical coverage. If the empirical coverage is lower than the nominal coverage, then the statistical model is underestimating the credible intervals of the observable. Conversely, a greater empirical coverage than the nominal coverage implies an overestimation of the uncertainty intervals. In this work, we compute empirical coverages across each category of observable used during calibration for both CHiRP and CHUQ, and present the average coverage across these observables. Empirical coverage was determined by whether an experimental datapoint fell within the posterior predictive distribution at a specified nominal coverage. Experimental variance, $\sigma_i^2$, was not added to posterior-predictive intervals when computing empirical coverage.

\section{Experimental Data}
\label{sec:data}

All experimental data used in this study were retrieved from the EXFOR data library \cite{otuka2014}. Data was retrieved for calibration using \texttt{exfortools} \cite{exfortools}. The experimental data shown in Sec. \ref{sec:propagation_results} was accessed using \texttt{EXFORTABLES} \cite{koning2026}. Data sets lacking reported uncertainties were excluded from the present work.

A total of 917 data points made up our calibration data corpus, of which 530 were elastic proton scattering differential cross sections or Rutherford ratio angular distributions (d$\sigma_{pp}$/d$\Omega$), 304 were elastic neutron scattering angular distributions (d$\sigma_{nn}$/d$\Omega$), and 83 were proton analyzing power angular distributions ($A_{y,pp}$). All incident nucleon energies had values between 10 MeV and 65 MeV. A breakdown of this data corpus is included as supplemental material to this document, as are all other displayed EXFOR data.

\section{\label{sec:results}Results}

\subsection{\label{sec:calibration_results}Optical Potential Calibration}

The parameter posterior distributions of a selection of calibrated parameters are shown in Figures \ref{fig:zr-real-corner} and \ref{fig:zr-imaginary-corner}. Fig. \ref{fig:zr-real-corner} displays a subset of parameters related to radii and real potential depths, and Fig. \ref{fig:zr-imaginary-corner} displays those related to the imaginary potential depths. Correlations are shown in the off-diagonal.

Fig. \ref{fig:zr-real-corner} shows an appreciable shift in the central $v_0$ parameter, which adopts a central value more in line with the original CH89 evaluation. This causes the central contribution to the potential to be weaker than in the CHUQ case. The isovector parameter $v_t$ is slightly more constrained compared with CHUQ, though a computed Kullback-Leibler divergence \cite{Kullbach1951} of 0.31 indicates only a modest difference. A notable shift is also observed in the $r_0$ parameter, which adopts a central value equal to the original CH89 value. There are also differences in the correlation between parameters for CHiRP and CHUQ; for example CHiRP displays a stronger correlations between $(v_0,r_0)$ than CHUQ.

Fig. \ref{fig:zr-imaginary-corner} shows the distribution of imaginary term parameters. We find that the primary parameters that influence the absorptive contribution are $w_{v0},w_{ve0}, w_{vew}$. In CHiRP these parameters lead to weaker interior absorption than in CHUQ, as can be seen in Fig. \ref{fig:imaginary_terms}. This figure presents the radial behavior of the two absorptive potential components at 10, 20, and 40 MeV beam energies undergoing proton scattering from $^{90}$Zr. We observe that the volume component, $W_v$, maintains a very low value compared with CHUQ, exhibiting a narrower band of values. In contrast, CHiRP predicts a higher surface contribution compared with CHUQ, while also predicting a narrower range of values. 
%The combined absorptive potential (lower panels) is lower in magnitude in CHiRP compared with CHUQ. 
It is also observed that uncertainty in the contribution increases somewhat with incident particle energy.

Table \ref{tab:parameters} shows the parameter values of CH89, CHUQ, and CHiRP. The final three rows in Table \ref{tab:parameters} display the values of the three model error fraction hyperparameters, corresponding to the three observable types represented by the calibration corpus. When compared with CHUQ, the CHiRP calibration results in lower error fractions across all three data categories. However, it should be emphasized that CHUQ is a global potential, whereas CHiRP was calibrated solely on zirconium data.

A selection of inner $68^{\text{th}}$ percentile credible intervals of $(p,p)$ elastic scattering predictive posterior distributions are shown in Fig. \ref{fig:zr_p_scattering}. All posterior predictive distributions were calculated according to the method described in Section \ref{sec:workflow}. CHiRP results are shown in blue, CHUQ in orange, and EXFOR data in black. In most cases it is observed that CHiRP provides a more accurate description of the EXFOR data, while also exhibiting narrower uncertainty bands. An exception to this is $(p,p)$ at 49.35 MeV  for $^{92,94}$Zr.

Calculations made with CHUQ consistently produce lower median cross sections than CHiRP. This is also observed in Fig. \ref{fig:zr_n_scattering}, showing the results of $(n,n)$ elastic scattering. Panel (a) shows results for scattering at 10 MeV, where CHiRP visibly improves on experimental data coverage compared to calculations with CHUQ. Panel (b) shows results at 24 MeV, where CHiRP does not cover $^{92,94}$Zr experimental data as uniformly as CHUQ, though CHiRP does reproduce the first diffraction peak for all targets more closely than CHUQ.

%CHiRP coverage of EXFOR data is superior to CHUQ's in all shown examples. Good coverage is also observed at 49.35 MeV (Figure \ref{fig:zr_p_scattering}b). However, at this energy calculations of $^{94}$Zr and $^{92}$Zr scattering with CHUQ displays better empirical coverage than CHiRP. No trend is apparent across our results which indicates whether a calculation with CHUQ will exhibit better coverage than with CHiRP.

Plots showing the empirical coverage averaged across respective observables are shown in the top row of Fig. \ref{fig:zr_ec}. These show that, for the higher confidence levels, CHiRP achieves improved coverage of data used in its calibration compared to CHUQ. Near ideal coverage is observed for neutron elastic scattering and $(p,p)$ analyzing power. Proton elastic scattering cross sections apparently display an overestimation of uncertainty compared with CHUQ, arguably a more desirable state of affairs in applied contexts, e.g. reactor design. The bottom row of Fig. \ref{fig:zr_ec} presents the distributions of the model error fraction parameters for CHiRP and CHUQ. This highlights that CHiRP achieves comparable or improved coverage to CHUQ while adopting significantly reduced estimates of model error,  therefore less model discrepancy noise was added to its predictions. It was found that choice of log-likelihood scaling factor, $\lambda$, strongly impacted the posterior distributions of $\beta_l$ --- our final selection was informed by inspection of $^{90}$Zr($n,tot$) predictions (see Appendix \ref{app:lambda} for details). One should also keep in mind in this comparison that the fractional error $\beta$ is defined differently for CHiRP and CHUQ.
%which implies CHiRP's improved model specificity.}

\subsection{\label{sec:extrapolations}Validation and Extrapolation}
Calculations were performed to validate CHiRP. These involved cross section predictions for targets not included in the calibration data. $^{96}$Zr data was omitted from this corpus, and results of proton scattering predictions on this target are shown in Fig. \ref{fig:zr_ext}. Significantly, this data was included in CHUQ's calibration corpus, however it can be seen that CHiRP improves on CHUQ's coverage.

In addition, calculations were executed on non-zirconium targets, and these also exhibit improved empirical coverage compared to CHUQ. For example, Fig. \ref{fig:yttrium_scattering} shows differential neutron scattering cross sections from $^{89}$Y. CHiRP calculations demonstrate excellent coverage of this data, more closely following the trend of data. Likewise, results of scattering from strontium isotopes (Fig. \ref{fig:sr_scattering}) show improved data coverage compared with CHUQ.

Lastly, Fig. \ref{fig:mo_scattering} presents scattering results from molybdenum isotopes. These also demonstrate CHiRP's improved coverage of experimental data compared with CHUQ.

\subsection{\label{sec:propagation_results}Propagating Uncertainty to Other Reactions}

We now propagate the CHiRP and CHUQ posterior distributions through {\sc talys} to compute the credible intervals for neutron-induced compound nuclear reaction observables. Specifically, we predict the inner 68$^{\text{th}}$ percentile credible intervals for $(n,n')$, $(n,\gamma)$, $(n,2n)$, and $(n,p)$ reactions on $^{90}$Zr and $^{89}$Y. $(n,tot)$ calculations performed with {\sc talys} are also presented. There are two important points to make concerning the results in this section: First, no experimental data shown here was included in the calibration of CHiRP\footnote{However, $(n,tot)$ results did inform our selection of adopted log-likelihood scaling factor, $\lambda$ (see Appendix \ref{app:lambda}).} (or CHUQ). Therefore, mismatch of the model with the data does not reflect on the quality of the calibration itself. Second, these observables depend strongly on other inputs which have significant uncertainties themselves, such as the nuclear level density model. Rather than focusing on the agreement of our predictions with the data, we focus instead on whether there are important differences in the predictions with CHiRP versus CHUQ, and the magnitude of the propagated uncertainty.

We also provide a crude estimate of the scale of uncertainties due to other HF model inputs aside from the optical potential. Following previous work \cite{pogliano2023,chalil2024}, we obtained results using each combination of nuclear level density, gamma-ray strength function, and pre-equilibrium model included in {\sc talys-2.0}. Each of the 240 {\sc talys} input combinations were adopted using the central values of both CHiRP and CHUQ. %The list of models used to obtain this data, as they are referred to in the {\sc talys} documentation, is shared in Appendix \ref{app:hf_inputs}.

\subsubsection{\label{sec:zr_prop}$n$ + $^{90}$Zr reactions}
\label{sec:zr}

{\sc talys} predictions of the $(n,tot)$ cross section as a function of energy for $^{90}$Zr using CHiRP (blue) and CHUQ (orange) are shown in Figure \ref{fig:zr_total}. The width of the credible intervals produced with CHiRP are slightly smaller compared to those produced by CHUQ and there is a slight difference in the predicted energy dependence.

%\IB{These predictions do not replicate the improved coverage compared to CHUQ that was seen in scattering results. However, CHiRP matches the trend of EXFOR near 10 MeV and above 25 MeV more closely than CHUQ.}

Various reactions affected by compound processes are shown in Fig. \ref{fig:zr_combined}. The solid bands (blue for CHiRP and orange for CHUQ) correspond to the optical model uncertainty, the focus of the current study. The hashed bands are discussed later.
Fig. \ref{fig:zr_combined}a presents the results of calculating inelastic scattering to the first metastable state in $^{90}$Zr. %Though this data is sparse, it can be seen that calculations with both OMPs fail to replicate the available data. 
As is common to each of the shown reaction channels, the median CHiRP calculation predicts lower cross sections than CHUQ. However, the magnitude of the uncertainty obtained with CHiRP is similar to that with CHUQ. We observe that the OMP appears to be the dominant source of uncertainty in this channel compared with other HF inputs below about 5 MeV.
%The inelastic scattering results show that initially, below roughly 5 MeV, HF calculations and their constituent model inputs have relatively little effect on the final cross section compared with OMP inputs, however subsequently the HF contribution increases until roughly 13 MeV where the OMP contribution to uncertainty shrinks compared with the impact of HF input.}

Figure \ref{fig:zr_combined}b shows the results for $(n,\gamma)$ reactions. The predictions for CHiRP are similar to CHUQ and, for both parameterizations, the uncertainties from the optical potential are very small. 
%This demonstrates that the distribution of parameters in both cases is too narrow to vary the probability of radiative capture substantially. We additionally observe that that difference in central values does not significantly effect the resultswhen changing the HF input models, which in both cases succeed in capturing the majority of experimental data in the energy region displayed.}

$(n,2n)$ results are displayed in Fig. \ref{fig:zr_combined}c. For this observable, the uncertainty from the optical model is significant
(a $1\sigma \approx 100$ mb throughout the displayed region). Also, $\sigma_{(n,2n)}$ with CHiRP differs significantly from those with CHUQ. %Results with CHiRP appear to successfully cover available experimental data around 15 MeV, however more significant tuning of the HF inputs would be required to also cover higher energy data. Conversely, CHUQ more closely follows the trend of EXFOR data with default TALYS settings, and considering the spread derived from CN inputs EXFOR remains covered.}
Similar results are seen for $(n,p)$ (Fig. \ref{fig:zr_combined}d).

In order to understand the relative importance of the optical model uncertainty compared to other inputs in the calculations, we estimated how the choice of these other inputs impact these cross sections. The hashed bands in Fig. \ref{fig:zr_combined} correspond to the full spread of results obtained by executing computations with each combination of nuclear level density, gamma-ray strength function, and pre-equilibrium model in {\sc talys-2.0}% (see Appendix \ref{app:hf_inputs} for complete list of models)
, using the central CHiRP and CHUQ parameter values. By considering the many options offered in {\sc talys} we are exploring for which observables the optical model uncertainty are likely to be significant. The hashed bands represent input sensitivity estimated in a crude manner, and do not correspond to a full Bayesian analysis as performed for the optical potential. It is most likely an overestimation and should not be added in quadrature with the optical model uncertainty.

The comparison of solid bands and hashed bands in Fig. \ref{fig:zr_combined} suggests  that optical model uncertainties are likely most important for the inelastic channel, and least important for radiative capture.

\subsubsection{\label{sec:y_prop}$n$ + $^{89}$Y reactions}
\label{sec:y}

Next we consider reactions on $^{89}$Y. The results for $\sigma_{(n,tot)}$  are shown in Fig. \ref{fig:y_total}. Similar behaviors are observed as those seen in the $^{90}$Zr case (Fig. \ref{fig:zr_total}).
%CHiRP clearly does not replicate the degree of coverage observed with CHUQ, instead consistently predicting higher median cross sections than CHUQ up to energies above 20 MeV. However, CHiRP does improve on CHUQ's coverage at these higher energies.}

In Figure \ref{fig:y_combined} we show the results for $\sigma_{(n,n')}$ to the first metastable state of $^{89}$Y, for $\sigma_{(n,\gamma)}$, for $\sigma_{(n,2n)}$, and for $\sigma_{(n,p)}$. We observe similar trends as the zirconium case, such as that the magnitude of cross sections from CHiRP are systematically smaller than CHUQ, but that the magnitude of the optical model uncertainty is similar. Finally, by comparing the range of cross sections obtained when varying HF input models, we conclude that the optical model uncertainty is most important for $\sigma_{(n,n')}$ and least important for $\sigma_{(n,\gamma)}$.

\section{\label{sec:conclusions}Conclusions}

In this work, we explore the concept of an uncertainty-quantified regional optical potential calibrated along a single isotopic chain, with the goal of improving predictions in specific regions of the nuclear chart when compared to predictions with global potentials. Our choice of potential form was guided by the fact that the Chapel Hill potential contains isoscalar and isovector components, which in principle should provide more reliable extrapolations of neutron/proton asymmetry to nearby nuclei. Results obtained with our potential, CHiRP, are systematically compared with those obtained with the global potential CHUQ.

We develop a statistical model that includes both experimental and model uncertainties, assuming a diagonal covariance matrix.  We perform a Bayesian calibration of the Chapel Hill parameters, using elastic scattering data on $^{90-94}$Zr isotopes, including both differential cross sections and polarization observables. Model uncertainties are introduced during calibration via a hyperparameter.

We find significant differences in the parameterization of CHiRP and CHUQ in the energy and radial dependence of their imaginary components. In particular, the volume imaginary term in CHiRP shows a reduced depth at low energies. Our results demonstrate that the regional potential, CHiRP, provides an improvement over the global potential, CHUQ, when considering elastic scattering data at energies around and above $E \approx 10$ MeV.

This improvement over CHUQ was also observed when extrapolating to nearby targets not included in our calibration data corpus. This indicates that regional potentials are a useful paradigm for predictive extrapolation to neighboring nuclei in data-poor regions important to applications.

We then use {\sc talys} to propagate the parametric uncertainties of both CHiRP and CHUQ to other reaction channels. We considered $(n,tot)$, $(n,n')$, $(n,\gamma)$, $(n,2n)$, and $(n,p)$. We find the cross sections predicted with CHiRP are generally lower than with CHUQ but the magnitude of the optical model uncertainty is similar. The comparison of the optical model uncertainty on these cross sections and a crude estimate of uncertainty due to other inputs on these same cross sections suggests that optical model uncertainties will be most important for the inelastic channel and least important for radiative capture. In particular, we observe that inelastic scattering is much more sensitive to the choice of optical potential than other Hauser-Feshbach inputs below roughly 5 MeV.

%We then use {\sc talys} to propagate the parametric uncertainties \IB{of both CHiRP and CHUQ} to other reaction channels. We considered $(n,tot)$, $(n,n')$, $(n,\gamma)$, $(n,2n)$, and $(n,p)$. \IB{We additionally present a crude estimate of uncertainty due to choice of nuclear level density, gamma strength function, and pre-equilibrium model. We observed the scale of uncertainty introduced by the OMP in these channels, as well as the effect of OMP selection on the HF calculations. The performance of both OMPs was mixed when considering reactions on $^{90}$Zr.%, and in particular the average results with CHUQ appear to better reproduce the $(n,2n)$ data. Strikingly however, when propagating to outside of the zirconium chain to $^{89}$Y, the average results of CHiRP followed the trend of experimental data better than CHUQ.} 

This work focuses on regional potentials in the near-spherical, middle-mass zirconium and yttrium region. Future work should expand this program to explore the optimal segmentation of the nuclear chart into regions for which such simple regional phenomenological optical potentials present excellent predictive power. 
%In addition, extrapolation to more exotic reactions and targets using our current potential is possible.

On the other hand, neither systematic uncertainties nor correlations in the model discrepancy across angle, energy, target, and reaction observable have been taken into account, in this study or past literature. Future work should explore ways to extend the statistical model used here to understand the impact of those effects.

\begin{acknowledgments}

Preliminary work was supported by the National Science Foundation CSSI program under award No. OAC-2004601 (BAND Collaboration \cite{phillips2021}). The work of F. M. N. was in part supported by the U.S. Department of Energy grant DE-SC0021422. This work relied on iCER and the High Performance Computing Center at Michigan State University for computational resources. S.~S. and P.~S. were supported in part by AWE Nuclear Security Technologies.  S.~S. was also supported by the UK EPSRC and P.~S. by the UK STFC under grant ST/Y000358/1. The authors would like to thank Dr.~Aaron Stott of AWE Nuclear Security Technologies for many invaluable discussions pertaining to this work. UK Ministry of Defence © Crown Owned Copyright 2026/AWE.

\end{acknowledgments}

\FloatBarrier

\appendix

%\section{\label{app:hf_inputs}Hauser-Feshbach Input Models}
%The models used to perform {\sc talys} calculations as described in Sec. \ref{sec:cn} are shared in Table \ref{tab:hf_inputs}.

\section{\label{app:chirp}CHiRP}

\begin{figure*}[!tbp]
    \centering
    \includegraphics[width=\linewidth]{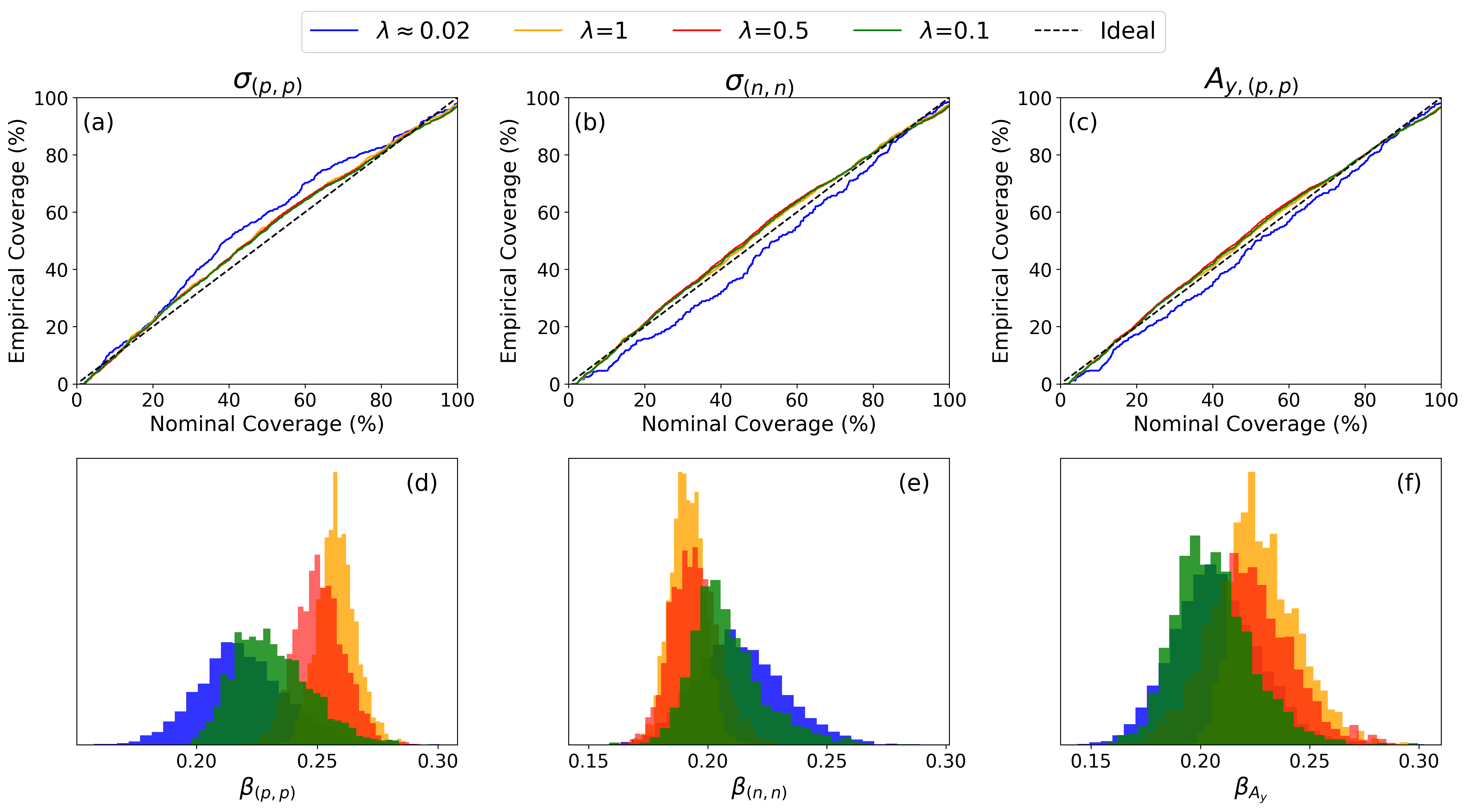}
    \caption{Panels (a), (b), and (c) show the average empirical coverage of CHiRP (blue) and calibrations using various scaling factor values, across all proton differential cross sections, neutron differential cross sections, and proton analyzing power data. Panels (d), (e), and (f) show the respective posterior distributions of the model error fraction parameters.}
    \label{fig:lambda_ec}
\end{figure*}

CHiRP is based on the same form as the CH89 OMP \cite{varner1991}, as is CHUQ. CH89 is a global OMP, meaning it was fitted using a corpus of experimental data obtained from a range of energies and across the nuclear chart; specifically, using elastic scattering cross sections and analyzing power datasets for incident energies in the range $E=10 - 65$ MeV, and target masses  $A=40 - 209$. This lower bound on scattering energy is significant, as below this value, significant compound contributions are observed.

CH89, CHUQ, and CHiRP consist of five terms:

\begin{equation}
\begin{split}
U(r, E)=&-\mathcal{V}_{r}(r,E)-i\Bigl(\mathcal{W}_v(r,E)+\mathcal{W}_s(r,E)\Bigr)\\
    &-\mathcal{V}_{so}(r)(\ell\cdot\sigma)+\mathcal{V}_C(r)
\end{split}
\end{equation}

These are the \textit{real} \textit{central} component ($\mathcal{V}_{r}$), \textit{imaginary} \textit{central} and \textit{imaginary} \textit{surface} components ($\mathcal{W}_v$ and $\mathcal{W}_s$), the \textit{real} \textit{spin-orbit} component ($\mathcal{V}_{so}$), and the Coulomb potential ($\mathcal{V}_{C}$). The Coulomb potential is zero in the case of non-charged particle reactions. The first three components have potential well \textit{depth} terms, and all but the Coulomb term have a Woods-Saxon or Woods-Saxon derivative \textit{radial} dependence. The components are defined as follows:

\begin{equation}
\begin{aligned}
\mathcal{V}_r(r, E) &= V_r(E) f(r,R_0,a_0)\\
\mathcal{W}_v(r, E) &= W_v(E)f(r,R_w,a_w) \\
\mathcal{W}_s(r, E) &= W_s(E)\times -4a_w\frac{d}{dr}f(r,R_w,a_w) \\
\mathcal{V}_{so}(r) &= 2V_{so}\times -\frac{1}{r}\frac{d}{dr}f(r,R_{so},a_{so}) \\
V_C(r) &= \begin{cases}
\dfrac{Zze^2}{2 R_C}\!\left( 3 - \dfrac{r^2}{R_C^2} \right) & r < R_C \\
\dfrac{Zze^2}{r} & r \ge R_C
\end{cases}
\end{aligned}
\end{equation}

In the Coulomb potential, $Z$ refers to the atomic number of the target nucleus, and $z$ is the atomic number of the projectile. Throughout, $R_x$ is a radius approximated as:

\begin{equation}
    R_x = r_xA^{1/3}+r_x^{(0)}
\end{equation}

\noindent where $A$ is the mass number of the target, and $r$ and $r^{(0)}$ are empirically fitted parameters; each radius parameter $R$ above is defined by a unique set of $r_x$ and $r_x^{(0)}$. The key difference between CHiRP and CHUQ is that CHiRP each intercept, $r_x^{(0)}$, is kept constant at its CH89 value. This is due to the fact that these relatively minor corrections were found to not be sufficiently constrained by the isotopic chain data. Therefore, CHiRP uses

\begin{equation}
R_x = \begin{cases}
 r_0A^{1/3}-0.225,& \text{for } R_0 \\
r_wA^{1/3}-0.42,& \text{for } R_w \\
r_{so}A^{1/3}-1.2,& \text{for } R_{so}\\
r_{c}A^{1/3}+0.12,& \text{for } R_{c}
.
\end{cases}
\end{equation}

The Woods-Saxon function is:
\begin{equation}
    f(r,R,a) = \frac{1}{1+e^{(r-R_x)/a_x} }
\end{equation}

\noindent where $a$ is the nuclear surface diffuseness parameter.

The potential well depths are defined as:

\begin{equation}
\begin{aligned}
V_r(E) &= v_0 +v_e\Delta E \pm a_{asym}v_t\\
W_v(E) &= w_{v0}\Bigl(1+e^{(w_{ve0}-\Delta E)/w_{vew}}\Bigr)^{-1} \\
W_s(E) &=(w_{s0}\pm a_{asym}w_{st})\Bigl(1+e^{(\Delta E-w_{se0})/w_{sew}}\Bigr)^{-1}. \\
\end{aligned}
\end{equation}

\noindent The asymmetry term, $a_{asym}$, is equal to ($N$-$Z$)/$A$), where $N$ is the neutron number of the target nucleus. The $\pm$ sign is positive for incident protons, and negative for reactions with neutrons. For $\Delta E$ we have:

\begin{equation}
    \Delta E = E - E_c
\end{equation}

\noindent Where $E_c$ is the volume-averaged Coulomb energy, which is non-zero for charged projectiles:

\begin{equation}
    E_c = \frac{6Zze^2}{5R_C}
\end{equation}

\noindent Otherwise, $E_c$ is zero.

For a complete list of parameters calibrated in this work and their evaluations, see Table \ref{tab:parameters}.

\section{\label{app:lambda}Likelihood Function Scaling}

In order to investigate the impact of the scaling factor $\lambda$, in addition to the $\lambda=k/N\approx0.02$ value ultimately adopted, calibrations were also obtained using values equal to 1, 0.5, and 0.1. The average empirical coverage across the three observation classes, as well as respective error fraction parameter posterior distributions, are shown in Figure \ref{fig:lambda_ec}.

This figure demonstrates that empirical coverage approaches the ideal line when the scaling factor is closer to 1, i.e. when the likelihood is not scaled at all. The three additional calibrations compared with the previously presented CHiRP calibration all demonstrate improved empirical coverage.

The effect on empirical coverage when changing the scaling factor from 1 to 0.1 is minimal, however the respective impact on the error parameter posterior distributions is more pronounced. In the case of the proton differential cross section error fractions in Figure \ref{fig:lambda_ec}d, the $\lambda=1$ results in convergence to a higher-value and narrower distribution compared with the $\lambda=k/N$ case, with each intermediate tested value progressively approaching the $k/N$ case. The situation is similar though less clear-cut in Figure \ref{fig:lambda_ec}f. Conversely, in the neutron scattering case we observe that the adopted error fractions are lower with $\lambda=1$, increasing as $\lambda$ is decreased ultimately to equal $k/N$. This implies that less scaling of the likelihood favors more accurate neutron scattering observables at the expense of the proton scattering observables. Separately, we also observe that convergence to a narrower posterior distribution when $\lambda$ is closer to one is also reflected in posterior predictive distributions of cross sections and analyzing power.

\begin{figure}[!htbp]
    \centering
\includegraphics[width=0.5\textwidth,keepaspectratio]{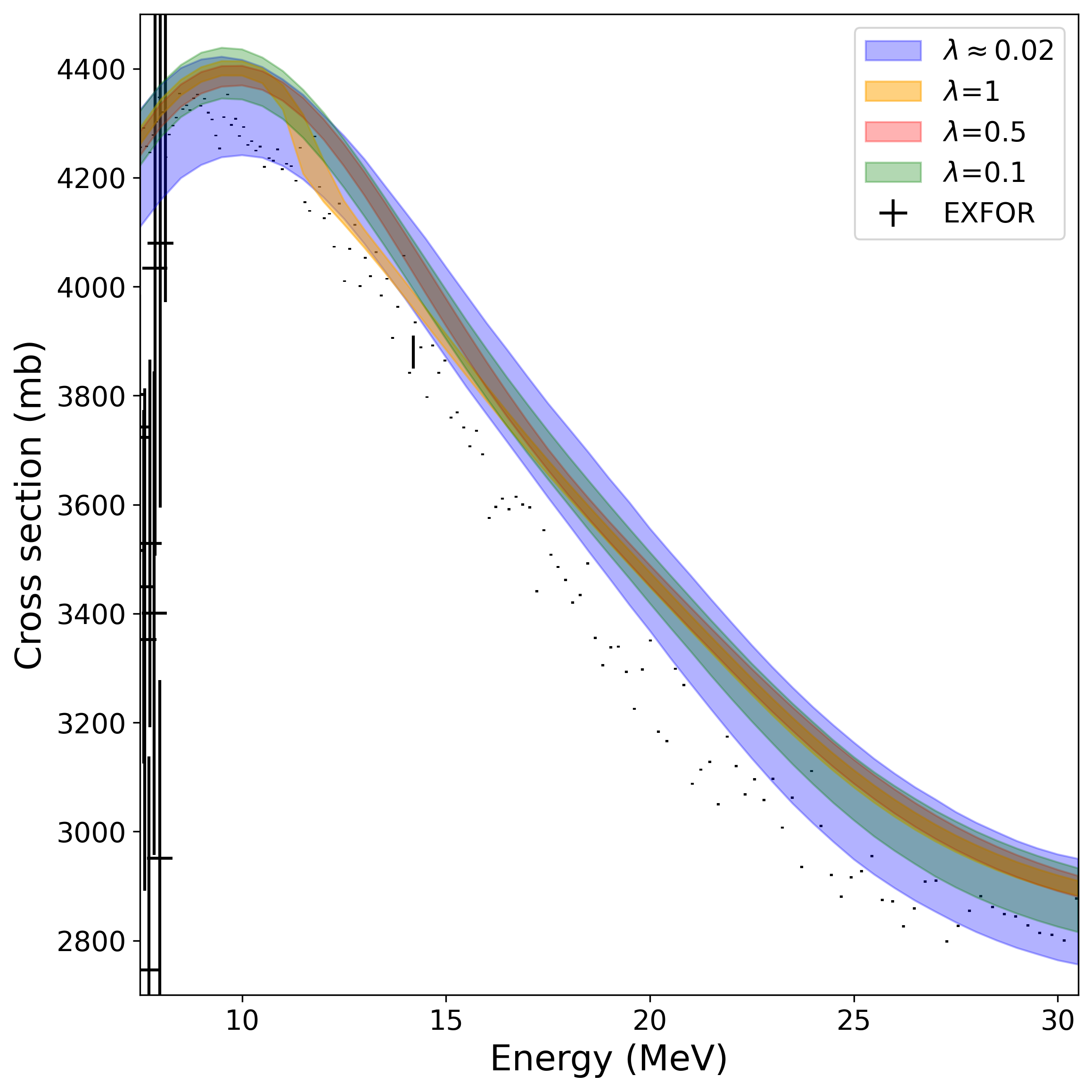}
    \caption{$^{90}$Zr$(n,tot)$ cross section results using CHiRP (blue) and various calibrations using alternative likelihood function scaling factors.}
    \label{fig:lambda_total}
\end{figure}

The systematic feature that weaker scaling results in narrower predictive distributions is particularly visible when inspecting calculated $(n,tot)$ cross sections, presented in Figure \ref{fig:lambda_total}. These results highlight that the $k/N$ results cover the available experimental data better than the alternatives. The observation that the $k/N$ results exhibit improved uncertainty bands compared with CHUQ, coupled with the best coverage of $(n,tot)$ data (which was not included in the calibration corpus), and a desire to balance the accuracy of predictions across the three calibration observable types, informed our decision to employ $k/N$ scaling in CHiRP.

\section{\label{app:nested_sampling}Nested Sampling}

\begin{figure}[!htbp]
    \centering

    \begin{subfigure}{0.45\textwidth}
        \centering
        \includegraphics[width=\textwidth,keepaspectratio]{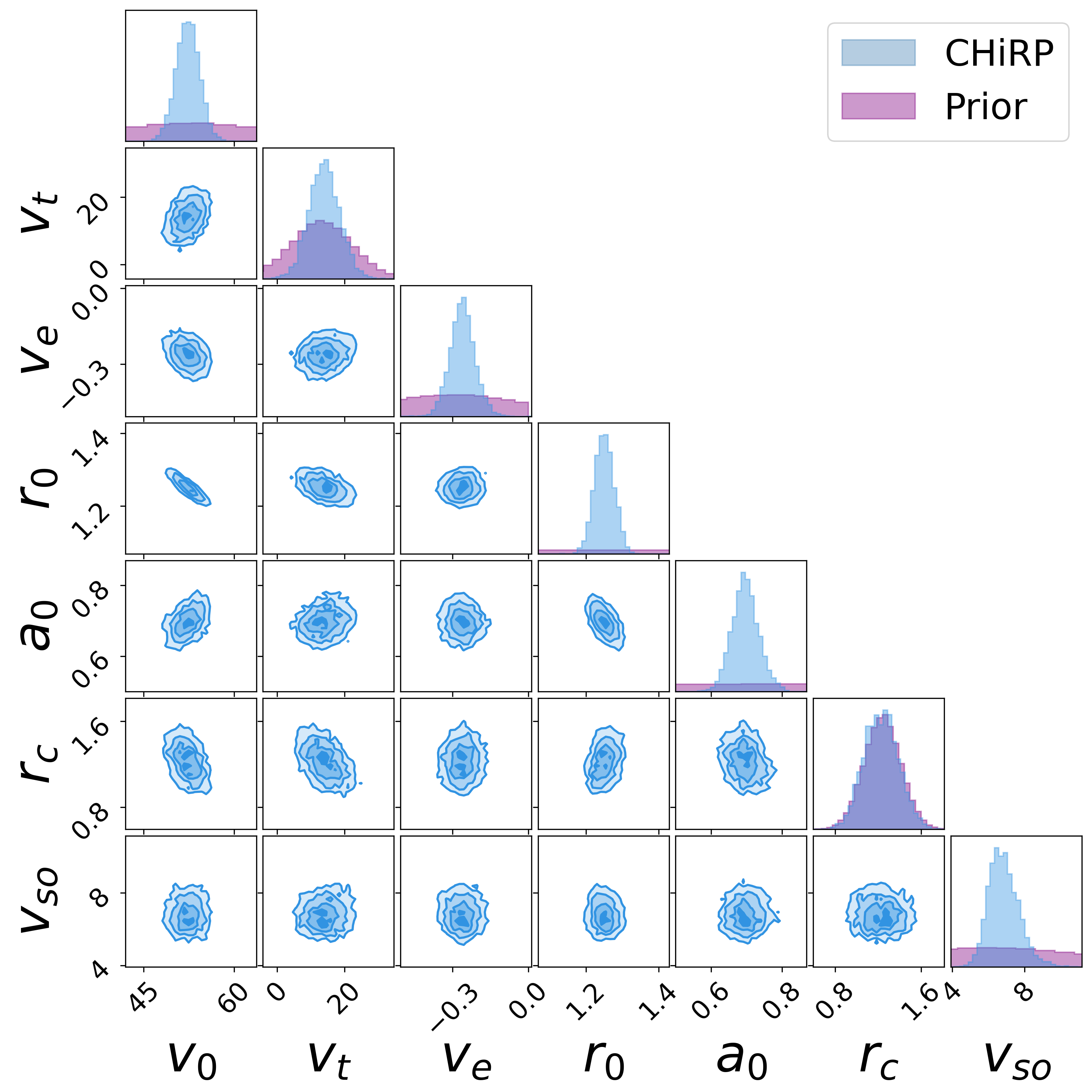}
        \caption{Real, central potential parameters and primary Coulomb radius parameter.}
        \label{fig:zr-real-corner-prior}
    \end{subfigure}

    \vspace{0.5em}

    \begin{subfigure}{0.45\textwidth}
        \centering
        \includegraphics[width=\textwidth,keepaspectratio]{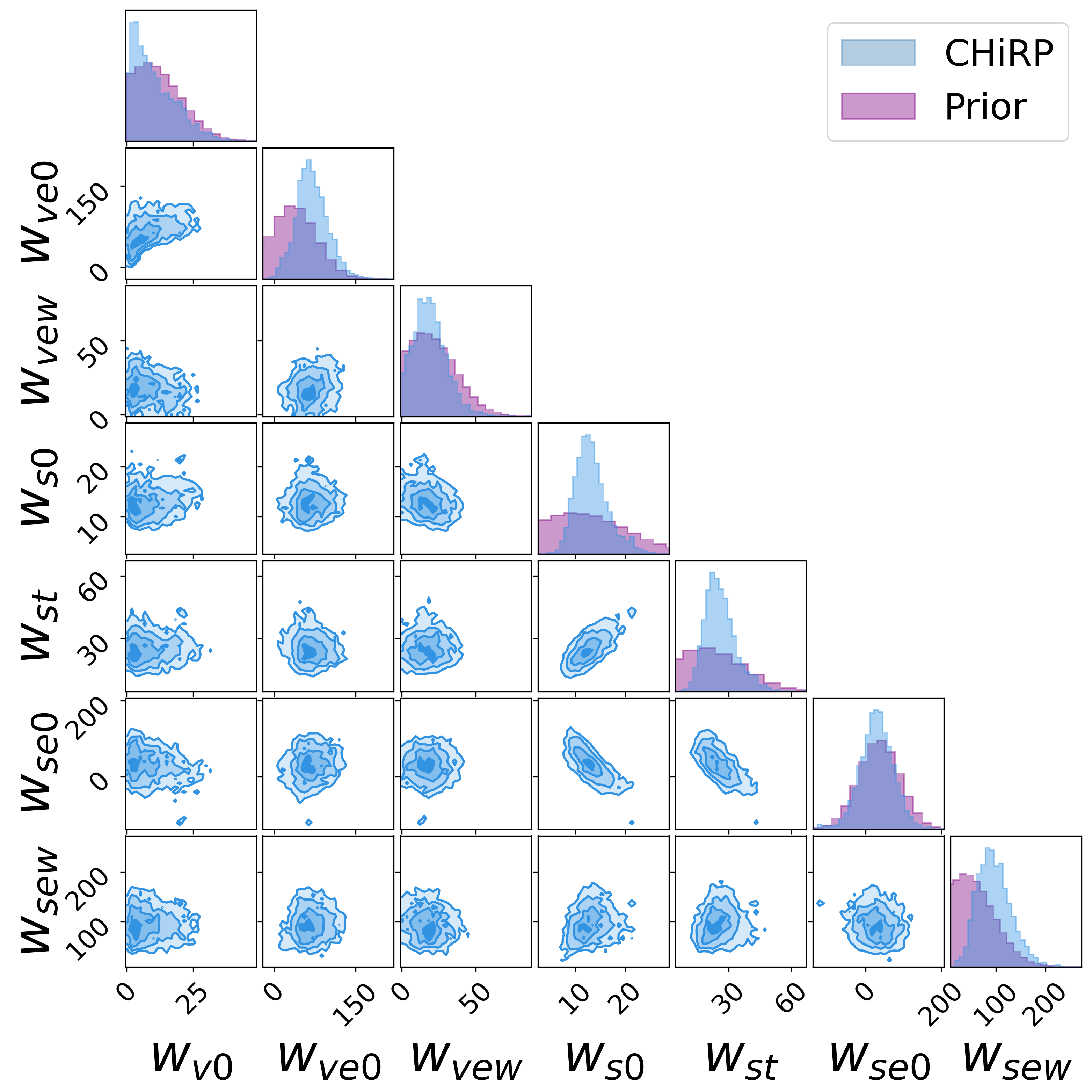}
        \caption{Imaginary potential parameters.}
        \label{fig:zr-imaginary-corner-prior}
    \end{subfigure}

    \caption{Corner plots displaying marginal posterior distributions (along the diagonal) and their correlations (in the off-diagonal). CHiRP results are shown in blue, CHiRP prior distributions are shown in purple.}
    \label{fig:zr-corner-combined-prior}
\end{figure}

\begin{table}[!htbp]
\centering
\begin{tabular}{|c|c|c|c|}
    \hline
    Parameter & Center & STD & STD \% \\
    \hline
    $v_0$ & 52.9 & 13.255 & 25 \\
    %\hline
    $v_t$ & 13.1 & 9.17 & 70 \\
    %\hline
    $v_e$ & -0.299 & 0.299 & 100 \\
    %\hline
    $r_0$ & 1.25 & 0.938 & 75 \\
    %\hline
    $a_0$ & 0.69 & 0.621 & 90  \\
    %\hline
    $w_{v0}$ & 7.8 & 11.7 & 150  \\
    %\hline
    $w_{ve0}$ & 35 & 42 & 120  \\
    %\hline
    $w_{vew}$ & 16 & 19.2 & 120  \\
    %\hline
    $w_{s0}$ & 10 & 10 & 100 \\
    %\hline
    $w_{st}$ & 18 & 18 & 100 \\
    %\hline
    $w_{se0}$ & 36 & 54 & 150 \\
    %\hline
    $w_{sew}$ & 37 & 55 & 150 \\
    %\hline
    $r_w$ & 1.33 & 0.998 & 75 \\
    %\hline
    $a_w$ & 0.69 & 0.518 & 75  \\
    %\hline
    $v_{so}$ & 5.9 & 5.9 & 100 \\
    %\hline
    $r_{so}$ & 1.34 & 1.005 & 75  \\
    %\hline
    $a_{so}$ & 0.63 & 0.473 & 75  \\
    %\hline
    $r_{c}$ & 1.24 & 0.17 & 13.71  \\

    \hline 
    $\beta_{pp}$ & 0.2 & 0.02 & 10 \\
    $\beta_{nn}$ & 0.2 & 0.02 & 10  \\
    $\beta_{Ay,pp}$ & 0.2 & 0.02 & 10 \\
    \hline
\end{tabular}
\caption{Configuration of CHiRP's multivariate Gaussian prior distribution. `STD' refers to the standard deviation of each Gaussian distribution, with `STD \%' referring to the standard deviation as a percentage of the central value.}
\label{tab:prior}
\end{table}

CHiRP was calibrated using static nested sampling \cite{skilling2006}. Nested sampling approximates the value of the Bayesian \textit{evidence}, the probability distribution describing the observed data, $p(\mathcal{D)}$. This is used to calculate our desired posterior distribution through Bayes' theorem \cite{sivia2006},

\begin{equation}
   p(\vars|\mathcal{D}) = \frac{p(\mathcal{D|\vars)}p(\vars)}{p(\mathcal{D)}},
\end{equation}

\noindent where the posterior is shown to be equal to the product of the likelihood and prior distributions (respectively shown on the numerator), and normalized by the evidence.

Nested sampling estimates $p(\mathcal{D})$ by constructing a prior volume by sampling a number of so-called `live' points from the prior distribution. From each of these prior samples, the point with the lowest likelihood is discarded and used to estimate the prior volume while a new live point is drawn. This procedure is then repeated. Each `dead' point has an associated weight, and the sum of the weights gives the approximated evidence evaluation. The posterior distribution is subsequently estimated as the ratio between weights of dead points and the approximated evidence.

We used \texttt{RxMC} to construct our calibration framework, which has integrated the Python package \texttt{dynesty} \cite{speagle2020} to perform nested sampling. Our sampler used a random walk algorithm and 500 live points. The stopping criterion, i.e. the estimated contribution from the unexplored prior volume, \texttt{dlogz}, was 0.5. The sampler was parallelized utilizing the University of Surrey's \textit{Eureka2} HPC nodes. Ultimately, 15,974 posterior samples were obtained.

Figure \ref{fig:zr-corner-combined-prior} shows corner plots of CHiRP in blue, with the prior distribution displayed in purple, and the marginal prior distributions' centroids in black. The same parameters as in Figure \ref{fig:zr-corner-combined} are presented. The central values and standard deviations (also expressed as percentages of the mean) used to define each dimension of the multivariate Gaussian prior distribution are presented in Table \ref{tab:prior}.

\FloatBarrier

\bibliography{references}

\end{document}